\documentstyle[aps,prb,psfig,multicol]{revtex}
\begin{document}

\newcommand{\uprule}{\end{multicols}}
\newcommand{\downrule}
{
\begin{multicols}{2}\narrowtext}
\draft
\sloppy

\title{Phenomenological description of the microwave
surface impedance and complex conductivity
of high-$T_c$ single crystals}

\author{M.~R.~Trunin and Yu.~A.~Nefyodov}
\address{Institute of Solid State Physics, 142432 Chernogolovka,
Moscow district, Russia}
\author{Herman~J.~Fink}
\address{Department of Electrical and Computer Engineering,
University of California, Davis,
California 95616, USA}
\date{\today}
\maketitle
\begin{abstract}
\noindent
Measurements of the microwave surface
impedance $Z_s(T)=R_s(T)+iX_s(T)$ and of the complex conductivity
$\sigma_s(T)$ of high-quality, high-$T_c$ single crystals of YBCO,
BSCCO, TBCCO, and TBCO are analyzed. Experimental data
of $Z_s(T)$ and $\sigma_s(T)$ are compared with calculations based on
a modified two-fluid model which includes temperature-dependent
quasiparticle scattering and a unique temperature variation of the
density of superconducting carriers. We elucidate agreement as well
as disagreement of our analysis with the salient features of the
experimental data. Existing microscopic
models are reviewed which are based on unconventional symmetry types of
the order parameter and on novel mechanisms of quasiparticle relaxation.
\end{abstract}

\vspace{0.15in}

\begin{multicols}{2}

\section{Introduction}

High-precision microwave measurements of the temperature dependence
of the surface impedance $Z_s(T)=R_s(T)+iX_s(T)$ of high-$T_c$
superconductors (HTS's) advance considerably our understanding of
pairing of superconducting electrons in these materials. In particular,
in 1993, the observed linear $T$--dependence of the penetration depth,
$\lambda(T)-\lambda(0) \propto \Delta X_s(T) \propto T$ below $25$ K
in the $ab$-plane of high quality YBa$_2$Cu$_3$O$_{6.95}$ (YBCO) single
crystals \cite{Har} gave rise to productive investigations of the order
parameter of HTS's. Such linear variation of $\lambda (T)$ at low $T$
has by now been observed not only in orthorhombic YBCO single crystals
\cite{Ach,Kam,Bon3,Mao,Bon33,Jac,Tru2,Srik1,Tru4,Srik2,Tru5,Kam8,Bon100}
and films \cite{Vau,Hens,Vau1,Efim}, but also in tetragonal
Bi$_2$Sr$_2$CaCu$_2$O$_8$ (BSCCO) \cite{Jac1,Shib,Lee,Tru90},
Tl$_2$Ba$_2$CuO$_{6+\delta }$ (TBCO) \cite{Broun,Waldr} and
Tl$_2$Ba$_2$CaCu$_2$O$_{8-\delta }$ (TBCCO) \cite{Tru4} single crystals.
This temperature dependence is not in accord with a nearly
isotropic superconducting gap and it is now considered to provide strong
evidence for $d$-wave pairing in these materials \cite
{Pin1,Pin2,Pin3,Hir1,Car1,Won,Hir2,Scal,Legg1,Maki,Izu},
in spite of the fact that the experimental data are not sensitive to
the phase of the superconducting
order parameter. Later research has shown that $\Delta \lambda _{ab}(T)$
could be linear at low $T$ for models invoking the proximity effect
between normal and superconducting layers \cite{Klem} or assuming
anisotropic $s$-wave pairing \cite{Kre,Zhu1,Adr}. However, none of these
theories can explain substantially different slopes of
$\Delta \lambda_{ab}(T)$ at low $T$ of YBCO samples grown by different
methods \cite{Tru9} nor features, such as a bump \cite{Srik1,Srik2,Hens,Nkl}
or a plateau \cite{Tru2,Tru4,Tru5}, observed in the intermediate temperature
range $0.3\,T_c<T<0.8\,T_c$.
Models containing a mixed $(d+s)$ symmetry of the order parameter \cite
{Joyn,Comb,Don1,Kim,Zaz1,Pok,Chub,Ren,Shap,Liu1,Pash,Beal1,Sch,Modr,YuraLT22}
hold some promise for a successful description of these experimental
features, but this would require additional theoretical investigations.

Another important feature of the microwave response of HTS crystals is
the linear variation with temperature of the surface resistance $R_s(T)$
in the $ab$-plane at low temperatures. At frequencies of about $10$~GHz
and below the $T$--dependence of $R_s(T)$ in BSCCO, TBCO, and TBCCO
single crystals is
linear over the range $0 < T \lesssim T_c/2$ \cite{Jac1,Lee,Tru90,Broun}.
For YBCO crystals $\Delta R_s(T)\propto T$ for $T\lesssim T_c/3$ and
$R_s(T)$ displays a broad peak and valley at higher temperatures
\cite{Bon3,Mao,Bon33,Jac,Tru2,Srik1,Tru4,Srik2,Tru5,Kam8,Bon100,Bon1,Bon2,Bon4,Kit,Tru100}.
This peak can be understood in terms of a
competition between an increase in the quasiparticle lifetime and a
decrease in the quasiparticle density as the temperature is lowered.
The fairly slow decrease in the
quasiparticle density is indicative of a highly anisotropic or
unconventional order parameter, resulting in a very small or vanishing
energy gap, while the increase in the quasiparticle lifetime is
attributed to the presence of inelastic scattering, which can be (i) due to
the exchange of antiferromagnetic spin fluctuations \cite{Bul}, which would
naturally lead to $d$-wave pairing, or (ii) due to strong electron-phonon
interaction \cite{El1,Pick1,Gin} within the anisotropic $s$-wave pairing
model \cite{Zhu2,Bil}. Moreover, there have been
suggestions of unconventional states for describing the charge carriers in
the CuO planes  like the marginal Fermi liquid \cite{Abra1,Abra2} and the
Luttinger liquid \cite{PW1,PW2}. However, to fit the data of YBCO, the
inelastic scattering rate has to decrease with temperature much faster than
any of these
microscopic models would predict. Further, the $d$-wave model, with point
scatterers, does predict a finite low temperature and low frequency limit,
which is independent of the  concentration and the strength of the
scattering centers \cite{Plee}. Therefore, the latter model does not explain
the very different values of the observed residual surface resistance
$R_{{\rm res}}\equiv R_s(T\to 0)$ on different samples. Furthermore, the
value of this universal surface
resistance is much lower than the $R_{{\rm res}}$--values obtained from
experiments. There is no microscopic theory which explains the linear
temperature dependence of $\Delta R_s(T)$ up to $T_c/2$ in the crystals
with non-orthorhombic structure and the  shoulder of $R_s(T)$ observed on
YBCO \cite{Srik1,Srik2} for $T>40$~K.

In the absence of a generally accepted microscopic theory a modified
two-fluid model for
calculating  $Z_s(T)$ in HTS single crystals has been proposed
independently in Refs.~\cite{Tru3,Fink1} and then further developed in
Refs.~\cite{Tru2,Tru9,Tru100,Fink2}. Our phenomenological model has
two essential features different from the well-known Gorter-Casimir model
\cite{Gor}. The first is the introduction of the temperature dependence of
the quasiparticle relaxation time $\tau(t)$ ($t\equiv T/T_c$) described by
the Gr\"uneisen formula (electron--phonon interaction), and the second
feature is the unique density of superconducting electrons $n_s(t)$ which
gives rise to a linear temperature dependence of
the penetration depth in the $\it ab$--plane at low temperatures

\begin{equation}
\lambda^2(0)/\lambda^2(t)=n_s(t)/n\simeq n(1-\alpha t),
\label{Llin}
\end{equation}

\noindent
where $n=n_s+n_n$ is the total carrier density, and $\alpha$ is a
numerical parameter in our model.

The goal of this paper is to demonstrate the power of our model
to describe the general and distinctive features of the surface
impedance $Z_s(T)$ and the complex conductivity $\sigma_s(T)$
in the superconducting
and normal states of different HTS crystals at various frequencies.
The following section describes the systematization
of the $Z_s(T)$ measurements, including the analysis which is used to
extract  $\sigma_s(T)$ from the measured values of $Z_s(T)$.
Section~III compares experimental data of $Z_s(T)$ and $\sigma_s(T)$
over the entire temperature range with calculations based on our modified
two-fluid model. In the conclusion we compare the concepts of our model
with results of microscopic theories. We hope that this will be a helpful
guide for future investigations of  microwave properties
of HTS's from a microscopic point of view.

\section{Analysis of experimental results}
\subsection{Surface impedance}

The surface impedance of HTS's, in terms of the complex conductivity
$\sigma_s=\sigma_1-i\sigma_2$, obeys the local equation (even at
temperatures $T\ll T_c$):

\begin{equation}
Z_s=R_s+iX_s=\left( \frac{i\omega \mu _0}{\sigma _1-i\sigma _2}\right)
^{1/2} \ . \label{Imp}
\end{equation}
The impedance components are

\begin{equation}  \label{RS}
R_s=\sqrt {{\frac{{\omega\mu_0(\varphi ^{1/2}-1)} }{{2\sigma _2\varphi }}}},
\end{equation}

\begin{equation}  \label{XS}
X_s=\sqrt {{\frac{{\omega\mu_0(\varphi ^{1/2}+1)} }{{2\sigma _2\varphi }}}},
\end{equation}
where $\varphi = 1+(\sigma _1/\sigma _2)^2$. It is obviously that $R_s<X_s$
for $T<T_c$.

For temperature $T<T_c$, if $\sigma_1\ll \sigma_2$, Eqs.~(\ref{RS}) and
(\ref{XS}) reduce to

\begin{eqnarray}
\label{R-X}
R_s  \simeq \ {\frac{{(\omega\mu_0)^{1/2}\sigma_1}}
{{2\sigma_2^{3/2}}}} & = & \frac{1}{2}\omega^2\mu_0^2\sigma_1\lambda^3,
\nonumber\\
X_s  \simeq  (\omega\mu_0/\sigma_2)^{1/2} & = & \omega\mu_0\lambda.
\end{eqnarray}

The components of the surface impedance are measurable quantities. The real
part of the surface impedance, the surface resistance $R_s$, is
proportional to the loss of the microwave power. It is caused by the presence
of ``normal'' carriers. In the centimeter wavelength band, typical
values of the surface resistance in the $ab$-plane of HTS single crystals
are between 0.1 and 0.3~$\Omega $ above but near the transition
temperature $T_c$. When $T$ is decreased through $T_c$, the surface
resistance abruptly drops, but does not seem to approach zero when $T\to 0$.
In conventional superconductors, like Nb, $R_s(T)$ decreases
exponentially with decreasing temperature below $T_c/2$, approaching a
constant residual surface resistance $R_{{\rm res}}$ as
$T\to 0$. $R_{{\rm res}}$ is due to the presence of various defects in the
surface layer of the superconductor . Therefore, it is generally accepted
that the lower the $R_{{\rm res}}$, the better the sample quality. In
high-quality HTS's there is no plateau in $R_s(T)$ at $T\ll T_c$. However,
we shall extrapolate the value of $R_s(T)$ to $T=0$~K and denote it by
$R_{{\rm res}}$. The origin of
the residual surface resistance observed in HTS crystals remains unclear.
It is known that $R_{{\rm res}}$ is strongly material and sample dependent
and is approximately proportional to the square of the frequency. At present,
very small values of $R_{{\rm res}}\sim 20$~$\mu\Omega$ at frequencies
$\sim 10$~GHz have been observed in YBCO single crystals \cite{Srik1,Bon100}.

The imaginary part of the surface impedance, the reactance $X_s$, is mainly
determined by the superconducting carriers and is due to nondissipative
energy stored in the surface layer of the superconductor.

In Table~I \cite{Tru10} we summarized the main features of the
temperature dependencies of the surface impedance of high-quality YBCO,
BSCCO, TBCO, and TBCCO single crystals whose residual surface resistance
in the $ab$-plane, $R_{{\rm res}}$, at frequency of $\sim 10$~GHz is less
than one milliohm, with $R_s(T_c)$ values of about 0.1~$\Omega $.
There is good reason to believe that the electrodynamic parameters of these
crystals adequately relate to the intrinsic microscopic properties of the
superconducting state of HTS.

To illustrate the data of Table I we show in Fig.~1, as an example,
experimental data of $R_s(T)$ and $X_s(T)$ in $ab$-plane of BSCCO single
crystal at 9.4~GHz \cite{Tru90}. In this figure $R_s(T)=X_s(T)$ for
$T\ge T_c$, which corresponds to the normal skin-effect condition.
Knowing $R_s(T_c)=\sqrt{\omega \mu_0 \rho(T_c)/2}\approx 0.12$~$\Omega$,
we obtain the resistivity $\rho(T_c)\approx 40$~$\mu\Omega\cdot$cm. In
the normal state, above $T_c$, the temperature dependence of $R_s(T)=X_s(T)$
is adequately described by
the expression $2R_s^2(T)/\omega \mu_0=\rho(T)=\rho_0+bT$. For the BSCCO
crystal in Fig.~1, $\rho_0\approx 13$~$\mu\Omega\cdot$cm and
$b\approx 0.3$~$\mu\Omega\cdot$cm/K.
The insets in Fig.~1 show $R_s(T)$ and $\lambda (T)=X_s(T)/\omega\mu_0$
for $T<0.7\,T_c$, plotted on a linear scale.
The extrapolation of the low-\linebreak

\uprule
\begin{table}
\caption{Surface impedance $Z_s(T)=R_s(T)+iX_s(T)$ in the $ab$-plane
of high-$T_c$ single crystals at frequencies $\sim 10$~GHz}

\vspace{3mm}
\begin{tabular}{c|c|c|c|c}
&\multicolumn{3}{c|}{Superconducting state, $T<T_c$}&Normal\\[2mm]
\cline{2-4}
HTS&Low temperatures&Intermediate temperatures&&state\\[2mm]
&4~K$<T\ll T_c$&$T\sim T_c/2$&$T\to T_c$&$1.5\,T_c>T\ge T_c$\\[2mm]
\cline{1-5}
Orthorhombic&$\Delta R_s(T)\propto T, \Delta X_s(T)\propto T$&Broad peak
in $R_s(T)$ at $25<T<45$~K [4-14]&&\\[2mm]
structure&at $T\lesssim T_c/4$;&\underline{Peculiarities:}\qquad 1. Shoulder
[9,11] in $R_s(T)$&Different&Normal\\[2mm]
YBCO&Essentially different slope&\quad at $T>40$~K;
\quad 2. Bump [9] or plateau [8,10]&slope of&skin-effect\\[2mm]
$T_c\approx 92~K$&of $\Delta\lambda(T)\propto T$ [1-14]&\qquad
on the curves of $X_s(T)$ at $50<T<80$~K&$\lambda(T)$ [3-14]&\\[2mm]
\cline{1-3}
Tetragonal structure&\multicolumn{2}{c|}{}&&\\[1mm]
BSCCO&\multicolumn{2}{c|}{}&Rapid&$R(T)=X(T)$\\[1mm]
$T_c\approx 90$~K [19-22]&\multicolumn{2}{c|}{$\Delta R_s(T) \propto T,
\qquad T\lesssim T_c/2$}&growth of&=\\[1mm]
TBCO&\multicolumn{2}{c|}{}&$R_s(T)$&$\sqrt{\omega\mu_0\rho(T)/2}$\\[1mm]
$T_c\approx 80$~K [23,24]&\multicolumn{2}{c|}{$\Delta X_s(T)=
\omega\mu_0\Delta\lambda (T)\propto T, \qquad T\lesssim T_c/3$}&and&\\[1mm]
TBCCO&\multicolumn{2}{c|}{}&$X_s(T)$&
$\Delta \rho(T)\propto T$\\[1mm]
$T_c \approx 110$~K [10,12]&\multicolumn{2}{c|}{}&&\\[1mm]
\end{tabular}
\end{table}
\downrule
\noindent
temperature sections of these curves to $T=0$~K
yields estimates of $R_{{\rm res}}=0.5$~m$\Omega$ and
$\lambda_{ab}(0)=2600$~\AA~for this crystal.

\begin{figure}
\centerline{\psfig{figure=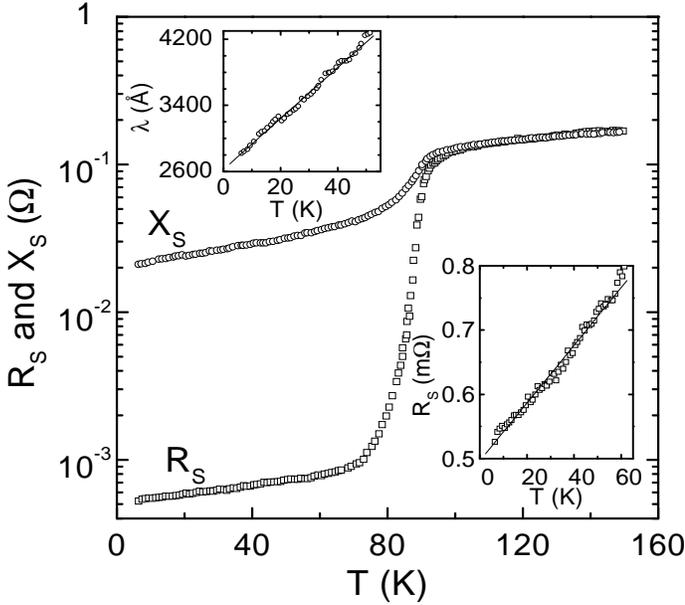,height=8cm,width=9cm,clip=,angle=0.}}
\caption{Surface resistance $R_s(T)$ and reactance $X_s(T)$ in $ab$-plane
of a BSCCO single crystal at $9.4$ GHz. The insets show linear plots
of $\lambda(T)$ and $R_s(T)$ at low temperatures. }
\label{r1}
\end{figure}

The experimental $\Delta\lambda_{ab} (T)$ of YBCO, TBCO, and TBCCO crystals
are also linear in the range $T<T_c/3$. It is important to notice the
different slopes of the $\Delta\lambda(T)\propto T$ curves for $T\ll T_c$. In
particular, in  YBCO crystals, fabricated by different techniques, the slopes
of $\Delta\lambda_{ab} (T)$
differs by almost one order of magnitude \cite{Tru2,Srik1,Kam8}. The reasons
for such a discrepancy are still unclear.

At frequencies of about $10$~GHz and below, the linear dependence $\Delta
R_s(T)\propto T$ in BSCCO (Fig.~1), TBCCO, and TBCO single crystals may
actually extend to temperatures of $\sim T_c/2$. This property, common for
all HTS crystals with the tetragonal structure, is not characteristic of
YBCO. As was noted previously, all microwave measurements on high-quality
YBCO single crystals show a broad peak in the $R_s(T)$ curve centered near
30--40~K up to frequencies of $\sim 10$~GHz. The peak shifts to
higher temperatures and diminishes in size as the frequency is increased.
In YBCO crystals of higher quality the amplitude of the peak increases and
$R_s(T)$ reaches its maximum at a lower temperature \cite{Bon100}.

The underlying origin of this YBCO feature has remained unclear. The
simplest idea is that the absence of
this peak in crystals with tetragonal structure might be caused by their
``poor'' quality, as is the case in YBCO doped with Zn \cite{Ach,Bon3,Bon2}.
However, this deduction is probably incorrect because, (i), there is a
sufficiently large set of experimental data indicating that $R_s(T)$ is a
linear function of $T$ for BSCCO,
TBCO, and TBCCO, and (ii), the peak in $R_s(T)$ was also detected in
such YBCO crystals \cite{Jac,Tru4,Kit} with parameters
$R_{{\rm res}}$ and $\rho(T_c)$  which would characterize the quality of
these crystal as ``poor'' compared to those of, for example, TBCCO \cite{Tru4}
or BSCCO \cite{Lee}. Results of the latter crystals are shown in Fig.~2.
The more probable cause of the peak, however, is the presence of an
additional component in the YBCO orthorhombic structure, namely CuO chains,
which lead to a mixed ($d+s$) symmetry of the order parameter in YBCO.
The electrons of the chains form an additional band, contributing to the
observed $T$--dependence of $Z_s(T)$. This contribution
seems to result in another\linebreak
\begin{figure}
\centerline{\psfig{figure=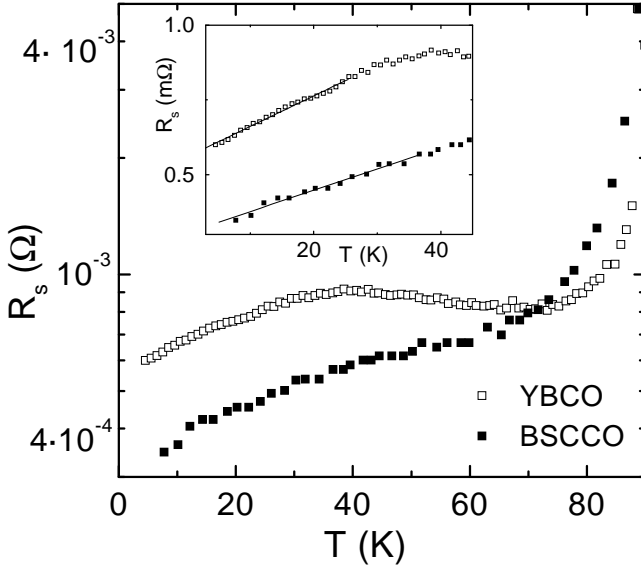,height=7.5cm,width=8.5cm,clip=,angle=0.}}
\caption{Comparison of the temperature dependencies of surface resistance
$R_s(T)$ of BSCCO and YBCO single crystals at 14.4~GHz.
Experimental data are taken from Refs.~\protect\cite{Lee} (BSCCO at 14.4~GHz)
and \protect\cite{Tru2} (YBCO at 9.4~GHz, scaled by $\omega^2$
to 14.4~GHz). The inset shows the linear $T$--dependencies of $R_s$ at
low $T$ for both materials, and a broad peak of $R_s(T)$ for YBCO.}
\label{r2}
\end{figure}
\noindent
distinctive feature of YBCO, namely a plateau
or bump (see Table I) on the $\lambda_{ab}(T)$ curve, which
has been observed in high-quality YBCO single crystals
\cite{Tru2,Srik1,Tru4,Srik2,Tru5} and films \cite{Hens,Nkl}.
However, recent measurements of $\Delta\lambda_{ab}(T)$ of YBCO
crystals,\cite{Bon100} grown in a high purity
BaZrO$_3$ crucible, show no such features in the
intermediate temperature range. The authors of Ref.\cite{Bon100} argue that
the disagreement with the results of Ref.\cite{Srik1} arises from some
problem connected with the surface of the crystal. The latter observation
still lack a convincing explanation.

Finally, another feature in the $T$--dependence of the impedance
of high-quality YBCO crystals was detected:
a noticeable increase of $R_s(T)$ with increasing temperature (shoulder)
at temperatures larger than the peak temperature at $30$~K.
It turns out that this shoulder was
reproducible in the experiments \cite{Srik1,Srik2}. Similarly, an explanation
of this observation is lacking.

\subsection{Complex conductivity}

Equations (\ref{Imp}--\ref{XS}) allows us to express the real and imaginary
parts of the complex conductivity
$\sigma_s=\sigma_1-i\sigma_2$ in terms of $R_s$ and $X_s$:
\begin{equation}
\sigma _1=\frac{2\omega \mu _0R_sX_s}{(R_s^2+X_s^2)^2},\qquad \sigma_2=
\frac{\omega \mu _0(X_s^2-R_s^2)}{(R_s^2+X_s^2)^2}.  \label{S-R}
\end{equation}

Above the superconducting transition temperature, the mean free path $\ell$
of current carriers is shorter than the skin depth $\delta_n$ in the normal
state (for $T\ge T_c$,\, $\ell\ll\delta_n$), which corresponds to the
conditions of the normal skin effect. Equations
(\ref{Imp}--\ref{XS}), (\ref{S-R}) also apply to the normal state of HTS's,
where $R_n(T)=X_n(T)=\sqrt{\omega\mu_0/2 \sigma_n(T)}$ with
$\sigma_n \equiv\sigma_1(T\ge T_c)$ and $\sigma_2 \ll \sigma_1$ at microwave
frequencies.

The components $\sigma_1(T)$ and $\sigma_2(T)$ are not measured directly
but derived from  measurements of $R_s(T)$ and $X_s(T)$ using
Eq.~(\ref{S-R}).

{\bf Low temperatures region ($T \ll T_c$)}

When $R_s(T)\ll X_s(T)$, then Eq.~(\ref{S-R}) reduces to:

\begin{equation}
\sigma _1(T)=\frac{2\omega \mu _0R_s(T)}{X_s^3(T)},\qquad \sigma _2(T)=
\frac{\omega \mu _0}{X_s^2(T)}.  \label{S12}
\end{equation}

It then follows from Eq.~(\ref{S12}), for low and intermediate
temperatures that $\sigma_1/\sigma_2=2R_s/X_s\ll 1$. The increments of
$\Delta\sigma_1(T)$ and $\Delta\sigma_2(T)$ depend on the increments of
 $\Delta R_s(T)$ and $\Delta X_s(T)$ relative to each other:

\begin{equation}
\Delta \sigma _1\propto \left(\frac{\Delta R_s}{R_s}-
3\frac{\Delta X_s}{X_s}\right), \qquad
\Delta\sigma _2\propto -\frac{\Delta X_s}{X_s}.  \label{DS12}
\end{equation}

It follows from Eq.~(\ref{DS12}) that the dominant changes of $\sigma_2(T)$
are determined mainly by the function $X_s(T)=\omega \mu _0\lambda(T)$,
reflecting the $T$--dependence of the magnetic field penetration depth.

The $T$--dependence of the real part of the conductivity, $\sigma _1(T)$,
is determined by the competition between the
increments $\Delta R_s/R_s$ and $\Delta X_s/X_s$.

In conventional
superconductors the quantity $X_s(T)$ ($\gg R_s$) is practically
$T$--independent ($\Delta X_s\approx 0$) at temperatures $T\le T_c/2$,
and $R_s(T)$ decreases exponentially, approaching the residual surface
resistance $R_{{\rm res}}$ as $T\to 0$. By subtracting $R_{{\rm res}}$ from
the measured $R_s(T)$, we obtain, using Eqs.~(\ref{S12}) and (\ref{DS12}),
the temperature dependence of $\sigma_1 (T)$ predicted by the BCS theory:
$\sigma_1=0$ at $T=0$, AND for $T\le T_c/2$, $\sigma_1(T)$ shows
an exponentially slow growth
with increasing temperature. Note that the smallest value
of $R_{{\rm res}}$ detected in pure Nb is, at least, two order of magnitude
smaller than the smallest value of
$R_{{\rm res}}$ measured in YBCO. The extremely
small values of the surface resistance in Eq.~(\ref{DS12}) indicate that
the increment $\Delta \sigma_1(T)$, in classical superconductors is always
positive  ($\Delta \sigma_1(T)>0$), at least in the temperature interval
$T<0.8\,T_{{\rm c}}$, before the maximum of BCS coherence peak is reached.

For HTS single crystals the $T$--dependence of $\sigma_1(T)$
is radically different from that predicted by theories of the microwave
response of conventional superconductors. In the $T$--range $T<T_c$ the
increments of $\Delta R_s(T)$ and $\Delta X_s(T)$ in HTS's are not small,
and, in addition, $\Delta X_s(T)\gg \Delta  R_s(T)$.
Although $R_s(T)<X_s(T)$, ${\Delta R_s}/{R_s}$ is not necessarily greater
than $3{\Delta X_s}/{X_s}$ in Eq.~(\ref{DS12}) or positive at all
temperatures. When that occurs, $\sigma_1(T)$ increases with
decreasing temperature. The function $\sigma_1(T)$ is maximum at some
$T=T_{{\rm max}}$, and then $\sigma_1(T)$ becomes smaller with
decreasing temperature. $\sigma_1(T)$ has a peak if the value of
$R_{{\rm res}}$ is sufficiently small when for $T\to 0$:

\begin{equation}
R_{{\rm res}}<\frac{X_s(0)}{3}\,\frac{\Delta R_s(T)}{\Delta X_s(T)}.
\label{Rres}
\end{equation}

If inequality (\ref{Rres}) is satisfied, $T_{{\rm max}}$ occurs at a finite
temperature, while for $R_{{\rm res}}$ being equal to the right hand side of
(9), $T_{{\rm max}}$ shifts to $0$ K. If $R_{{\rm res}}$ is such that (9)
is not satisfied, $\sigma_1(T)$ decreases at low temperatures as the
temperature is increased, which is quite different from what is observed
with conventional superconductors.

Thus, the shape of the $\sigma_1(T)$ for $T\ll T_c$ depends on the
value of the residual surface resistance $R_{{\rm res}}$, whose origin and
accurate value are unknown. For this reason, the shapes of $\sigma_1(T)$
curves are not determined unambiguously for $T\le T_c/2$, unlike the
functions $R_s(T)$ and $X_s(T)$, which are directly measured in
experiments.

If we linearly extrapolate $R_s(T)$ to $T=0$
and attribute the resulting $R_s(0)$ to the residual surface resistance,
$R_s(0)=R_{{\rm res}}$, and then substitute the temperature dependent
difference $R_s(T)-R_{{\rm res}}$ into the numerator of the first expression
of Eq.~(\ref{S12}), the result is that the  $\sigma_1(T)$ curve has
a broad peak for HTS materials. Near $T = 0$, $\sigma_1(T)$ increases
linearly with $T$ from zero, reaches a maximum at $T_{{\rm max}}$, and then
decreases to $\sigma (T_c)$. This procedure, however, ignores the
possibility of intrinsic residual losses. Therefore, some authors
(see, e.g., Refs.~\cite{Bon100,Lee,Bon4}) associate residual losses in
HTS single crystals with a residual normal electron fluid.
This implies that the source of the residual loss is in the bulk of the
sample, although it is probably not intrinsic. If this contribution is
excluded from the complex conductivity of the superconductor, one obtains
$\sigma_1(T=0)\to 0$, as can be seen in Fig.~3 from the measurements taken
at 13.4, 22.7, and 75.3~GHz
by the authors of Ref.~\cite{Bon100}. The peak of $\sigma_1(T)$ shifts to
higher temperatures and diminishes in size as the experimental frequency is
increased. In YBCO single crystals the temperature $T_{{\rm max}}$ at which
the maximum of $\sigma_1$ occurs is close to the temperature at which the
peak of $R_s(T)$ occurs.
\begin{figure}
\centerline{\psfig{figure=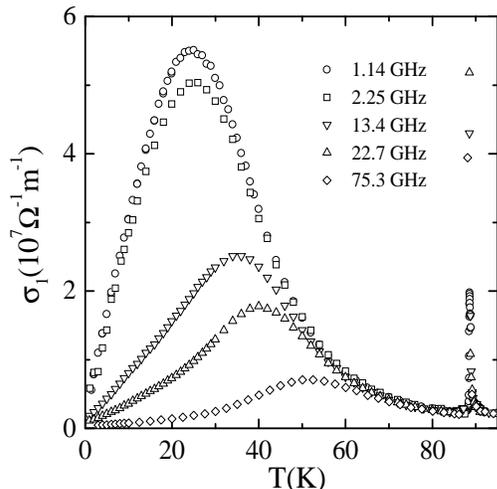,height=6.5cm,width=6.5cm,clip=,angle=0.}}
\caption{Real part of the conductivity $\sigma_1(T)$ of YBCO single crystal
at different frequencies \protect\cite{Bon100}. The data were obtained
courtesy of the Vancouver group (D.~A.~Bonn). }
\label{r3}
\end{figure}

Finally, one may procure $\sigma_1(T)$ from measurements of $R_s(T)$ and
$X_s(T)$ for $T>0$ without any concern about $R_{{\rm res}}$. In this case,
$\sigma_1(0)$ is not determined uniquely. Whether $\sigma _1(T)$ has a
peak or not depends on the validity of condition~(\ref{Rres}). The curves
at 1 and 2 GHz in Fig.~3 have been obtained using Eq.~(\ref{S-R}) without
subtracting any residual losses.

{\bf Temperatures close to $T_c$ ($T\to T_c$)}

Equations (\ref{S12}) and (\ref{DS12}) do not apply near $T_c$.
In this temperature range it is necessary to use the general local
relationships~(\ref{Imp}--\ref{XS}), (\ref{S-R}).

The conductivity $\sigma_2(T)$ in the $ab$-plane of HTS crystals abruptly
drops to very small values in the normal state. The expression
$(T_c/\sigma _2(0))d\sigma_2(T)/dT$ at $T=T_c$, defining the slope of
$\lambda^2(0)/\lambda^2(T)$ at $T=T_c$, varies between $-2$ and $-4$ for
different crystals.

The real part of the conductivity, $\sigma_1(T)$, does not show a coherence
peak near $T=0.85\,T_c$, as predicted by BCS. Usually, the real part of the
conductivity, $\sigma_1(T)$, of HTS single crystals has a narrow peak near
$T_c$ which increases with decreasing frequency \cite{Lee,Broun,Waldr}.
The width of the narrow peak of $\sigma_1(T)$ coincides with the
width of the $R_s(T)$  transition at microwave frequencies. A possible
explanation of the sharp peak just below $T_c$ is inhomogeneous
broadening of the superconducting transition \cite{Glas,Ols,Gol1} or
fluctuation effects \cite{Waldr,Horb,Anl1}.

\section{Modified two-fluid model}

As was shown in Ref.~\cite{Gin}, high $T_c$--values  ($T_c\sim 100$~K),
the temperature dependence of the resistivity, the frequency dependence of
the momentum relaxation time, and other properties of the normal state
in optimally doped HTS's are well described within the framework of the
Fermi-liquid approach, including strong electron-phonon coupling (SC)
\cite{El1}. The SC model also explains some of the features of the
superconducting state of HTS's.
It follows from the Eliashberg theory that the distinctive component of
superconductors with strong coupling is that the gap in the spectrum of
electronic excitations is smeared. Strictly speaking, there is no gap,
whatsoever, at $T\ne 0$ \cite{El3,Mak1}. This leads to breaking of Cooper
pairs, smearing of the peak in the density of states at
$\hbar \omega =\Delta(T)$ due to inelastic scattering of electrons by
thermally excited phonons, and suppression of coherence effects. As a
result, the amplitude of the coherence peak decreases and, according to
Refs.~\cite{Mar,Mak2}, virtually disappears at frequencies around 10~GHz
if the electron-phonon coupling constant exceeds unity. Moreover, the
mechanism of quasiparticle generation is radically different from that
of the BCS model. The quasiparticles are generated without jumps across the
energy gap and can be in states with all energies down to $\hbar \omega=0$.
These states can be classified as gapless, and the quasiparticles can be
treated \cite{Gin} as normal current carriers in the two-fluid model. So it
is not surprising that an important consequence of the SC model is the
nonexponential behavior of $R_s(T)$ \cite{Dol1} and $\lambda (T)$
\cite{El2}. Power-law temperature dependencies were also predicted by the
two-fluid Gorter-Casimir (GC) model \cite{Gor}, and near $T_c$ they proved
to be quite close to calculations performed by the SC model. In particular,
the curves of $\lambda^2(0)/\lambda^2(T)$, calculated by the SC
model, \cite{Kar1,Coll,Ram,Andre} proved to be fairly close to the function
$n_s(t)/n=1-n_n(t)/n=1-t^4$ in the GC model. The slopes of these curves
at $T=T_c$ are in agreement with those measured with different YBCO single
crystals and are equal to \cite{Bon3} $-3$  or \cite{Mao,Tru2,Tru4} $-4$.
The experimental fact that there is no BCS coherence peak
in the conductivity of HTS crystals, indicates the necessity of taking
into account strong coupling effects near $T_c$ and the feasibility of
interpreting HTS properties at microwave frequencies in terms of a
two-fluid model.

The complex conductivity $\sigma_s$ is a basic property of superconductors.
In accordance with GC model \cite{Gor} the expressions for the components
of $\sigma_s=\sigma_1-i\sigma_2$ are:

\begin{eqnarray}
\label{S1-S2}
\sigma _1 & = & {\frac{{n_ne^2\tau }}m}\left[ \frac{{1}}
                   {{1+(\omega \tau )^2}}\right], \nonumber\\
\sigma _2 & = & {\frac{{n_se^2}}{{m\,\omega }}}\left[ 1+{\frac{{n_n}}
           {{n_s}}}{\frac{{(\omega \tau )^2}}{{1+(\omega \tau )^2}}}\right].
\end{eqnarray}

At  temperatures $T\leq T_c$ the
total carrier concentration is $n=n_s+n_n,$  where $n_{s,n}$ are the
fractions of superconducting and normal carrier  densities (both have the
same charge $e$ and effective mass $m$).
The real part of conductivity $\sigma_1$ is determined purely by the normal
component, while both components, normal and superconducting,
contribute to the imaginary part $\sigma_2$.
The relaxation time $\tau$ of normal carriers in the GC model is
independent of temperature. This is acceptable if we assume that the
behavior of normal carriers in superconductors is similar to that of
normal carriers in normal metals at low temperatures.
Scattering of electrons at very low temperatures is due to impurities and
independent of the temperature. Therefore, the temperature dependence of
the real part of the conductivity (\ref{S1-S2}) in the GC model is determined
entirely by the function $n_n(T)$ with $n_s(T)=n-n_n(T)$ only.

For sufficiently low frequencies $(\omega \tau )^2\ll 1$ the expressions
of the conductivity components of Eq.~(\ref{S1-S2}) transform into
simple relations

\begin{equation}
\sigma _1={\frac{{e^2\tau } }{m}}n_n, \qquad \sigma _2={\frac{{e^2} }{{\
m\omega }}}n_s= \frac 1{\mu_0\omega\lambda^2},
\label{Sigma1}
\end{equation}
where $\lambda=\sqrt{m/\mu_0n_se^2}$ is the London penetration depth of
a static magnetic field.

Penetration of alternating fields into superconductors is controlled by
the frequency-dependent skin depth. Based on results of the complex
conductivity (\ref{Sigma1}), one obtains the complex skin depth $\delta_s$
by generalizing the corresponding expression for a normal conductor:

\begin{equation}  \label{Skin}
\delta_s=\frac{\sqrt{2}\lambda}{\sqrt{\omega \tau (n_n/n_s)-i}}.
\end{equation}

With increasing angular frequency $\omega$ the skin depth Re($\delta_s$)
decreases and, therefore, the London penetration depth $\lambda$
gives the upper bound for the penetration of the electromagnetic field into
a superconductor. In GC model the  $\lambda$ value diverges near $T_c$
as $\lambda(t)=\lambda/[2\sqrt{1-t}]$ and the function
$\sigma_2(t)/\sigma_2(0)=4(1-t)$ tends linearly to zero at $T=T_c$
with a slope equal to -4. At the same time, at $T=T_c$ the skin depth
Re($\delta_s$), defined by Eq.~(\ref{Skin}), crosses over to the skin depth
$\delta_n$ for a normal conductor.

\subsection{Scattering and surface resistance of HTS single crystals}

In conventional superconductivity one assumes that below $T_c$ the mean free
path does not vary with temperature. In a normal metal, at much higher
temperatures than the corresponding $T_c$ of a conventional superconductor,
the electron scattering rate is proportional to $T$~~\cite{Maks10}. Since the
transition temperatures of HTS's are much larger than those of conventional
superconductors, it stands to reason that temperature will affect the
electron scattering rate of the quasiparticles of HTS's below $T_c$,
but be limited to a constant rate at low temperatures. Therefore, if
a two-fluid model is to be successful in explaining transport properties
of HTS's, then it is natural to include a temperature variation of $\tau$
into that model.

To obtain an expression for $\tau(T)$, we rely on the
analogy between the `normal fluid' component in the superconducting state
and charge carriers in a normal metal.
According to Mathissen's rule, the reciprocal relaxation time at
temperatures below the Debye temperature $\Theta$ is

\begin{equation}
\frac{1}{\tau}=\frac{1}{\tau_{{\rm imp}}}+\frac{1}{\tau_{{\rm e-ph}}}.
\label{Tau}
\end{equation}

The first term on the right is due to impurity scattering and is a constant
of temperature, and the second is due to electron-phonon scattering and is
proportional to $T^5$.

From Eq.~(\ref{Tau}) we express $\tau(T)$ as
\begin{equation}
{\frac{1}{{\tau (t)}}}={\frac{1}{{\tau (T_c)}}}\,{\frac{{\beta
+t^5} }{{\beta +1}}},
\label{Tau1}
\end{equation}

\noindent
where $\beta$ is a numerical parameter: $\beta \approx \tau(T_c)/\tau(0)$,
provided this ratio is much less than unity. It should be pointed out,
however, that this approximation is not always satisfied.

Equation~(\ref{Tau1}) corresponds to the low-temperature limit of the
Bloch-Gr\"uneisen formula, which includes impurity scattering and can
be presented over a wide temperature range by the expression

\begin{eqnarray}
\label{Tau2}
\frac{1}{\tau (t)} & = & \frac{1}{\tau (T_c)}\,
\frac{\beta +t^5{\cal J}_5(\kappa/t)/{\cal J}_5(\kappa )}{\beta +1} \nonumber\\
{\cal J}_5(\kappa/t) & = & \int\limits_0^{\kappa /t}
\frac{z^5e^zdz }{(e^{z}-1)^2} \ ,
\end{eqnarray}

\noindent
where $\kappa =\Theta /T_c$. For $T<\Theta/10$ ($\kappa>10t$), Eq.~(\ref
{Tau2}) approaches the form of Eq.~(\ref{Tau1}). For $T>\Theta/5$ ($\kappa<5t$),
we obtain from Eq.~(\ref{Tau2}) the linear $T$--dependence of
$1/\tau(t)\propto t$.
Examples of $1/\tau(t)$  for different parameters of $\beta$,
$\kappa$, and $\tau(T_c)$ are shown in Fig.~4.
\begin{figure}
\centerline{\psfig{figure=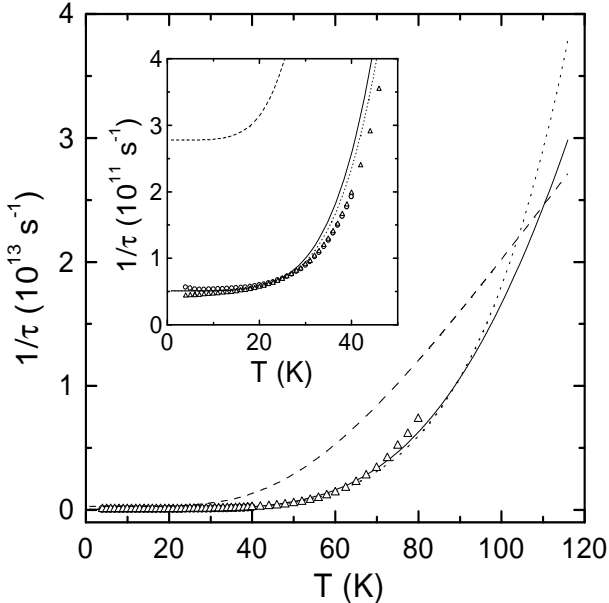,height=8cm,width=8cm,clip=,angle=0.}}
\caption{Scattering rate of quasi-particles, calculated from
Eq.~(\protect\ref{Tau1}), dotted line: $\beta=0.005$, and
Eq.~(\protect\ref{Tau2}), solid line: $\beta=0.005$, $\kappa=9$; dashed
line: $\beta=0.02$, $\kappa=4$. The triangles are calculated from
 $1/\tau=[1-\lambda^2(0)/\lambda^2(T)]/[\mu_o\sigma_1(T)\lambda^2(0)]$,
with $\sigma_1(T)$ and $\lambda(T)$ at 1.14 GHz and
$\lambda(0)=1600$~\AA~in the $ab$-plane, with currents parallel to
the $a$-direction of the YBCO crystal \protect\cite{Bon100}.
The inset shows the low temperature parts of the curves.
The circles are from Fig.~8 of Ref.~\protect\cite{Bon100}. }
\label{r4}
\end{figure}

For $\omega \tau(T_c) \ll 1$, which is normally satisfied at
microwave frequencies, the parameter $\omega \tau(T_c)$ is obtained from
measurements of $R_s(T_c)$ and $X_s(0)$:

\begin{equation}
\omega \tau(T_c)=
\frac{X_s^2(0)}{2R_s^2(T_c)}=
\frac{\sigma_1(T_c)}{\sigma_2(0)}.
\label{Omtc}
\end{equation}

At frequencies $\sim 10$~GHz, the value $\omega \tau$ of the best HTS
crystals is of the order of $10^{-3}$ at $T=T_c$ and remains less than
unity at all temperatures $T<T_c$, as will be discussed below. Therefore,
the expressions of the conductivity components in Eq.~(\ref{S1-S2})
in the two-fluid model turn into the simple form (\ref{Sigma1}).

All experimental data of $R_s(T)$ of high-quality HTS single crystals
can be elucidated by our two-fluid model with $\tau(T)$ given by
Eqs.~(\ref{Tau1}) or (\ref{Tau2}).

Measurements of $R_s(T)$ of YBCO single crystals
at frequencies of order or less than 10~GHz are analyzed first.
Values of $\sigma_2(T)/\sigma_2(0)=\lambda^2(0)/\lambda^2(T)=n_s(T)/n$,
measured in the same experiments, and $\sigma_1 (T)/\sigma (T_c)$ obtained
from Eq.~(\ref{Sigma1}) are substituted into the Eq.~(\ref{RS}).
Use is made of the relation $n_n(T)/n=1-\sigma_2(T)/\sigma_2(0)$, which is
obtained from the experimental data, and $\tau(T)$, employing
Eqs.~(\ref{Tau1}) or (\ref{Tau2}).

Setting $\beta=0.005$ and $\kappa=9$ in Eq.~(\ref{Tau2}) and taking the
experimental values $\sigma_2(T)/\sigma_2(0)$ from Fig.~11
(see below) and $\omega\tau (T_c)=7.5\times 10^{-4}$ at 1.14~GHz we
find from Eqs.~(\ref{Sigma1}) and (\ref{RS}) the  $T$--dependencies of
$R_s(T)$, shown by the curves in Fig.~5. These curves match the data of
Ref.\cite{Bon100} over the entire temperature range. The same result is
obtained using Eq.~(\ref{Tau1}) instead of Eq.~(\ref{Tau2}), with
$\beta=0.005$. For at $\kappa\gg 1$ and $T\alt T_c$,
Eqs.~(\ref{Tau1}) and (\ref{Tau2}) are identical.
\begin{figure}
\centerline{\psfig{figure=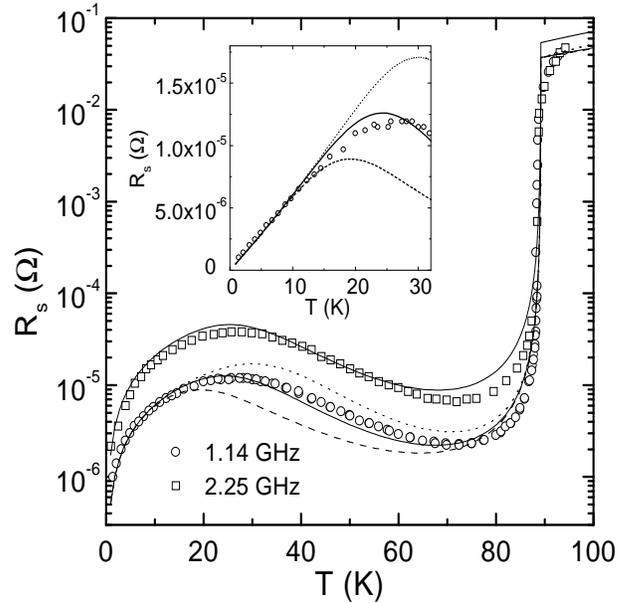,height=8cm,width=8cm,clip=,angle=0.}}
\caption{Experimental $R_s(T)$ data of YBCO single crystal
\protect\cite{Bon100} at 1.14~GHz (circles) and 2.25~GHz
(squares). Solid curves are calculations using Eqs.~(\protect\ref{RS}),
(\protect\ref{Sigma1}) and (\protect\ref{Tau1}). The dashed curves are
calculated at 1.14~GHz with the term $t^5$ replaced by $t^4$ in the
numerator of Eq.~(\protect\ref{Tau1}), the dotted curves with $t^6$.
The inset shows a linear plot of $R_s(T)$ at low temperatures at 1.14~GHz. }
\label{r5}
\end{figure}

From Eqs.~(\ref{R-X}) and (\ref{Sigma1}) it follows that for
$\alpha\,t\ll 1$ [see Eq.~(\ref{Llin})] a rough estimate of the temperature
$t_m$ at which $R_s(T)$ is maximum is obtained from the relation
$\beta\simeq 4\,t_m^5$. The value of $\tau(0)$ is found from the slopes
${dR_s}/{dT}$ and ${d\lambda}/{dT}$ of the experimental data of $R_s(T)$ and
$\lambda(T)$ as $T\to 0$ [$\omega \tau(0)<1$]:

\begin{equation}
\omega \tau(0)=
{\frac{1}{\mu_0\,\omega}}\,{\frac{dR_s}{d\lambda}}.
\label{Omt0}
\end{equation}

With Eq.~(\ref{Omtc}) and (\ref{Omt0}) the parameter
$\beta \approx \tau(T_c)/\tau(0)$ is determined from the surface
impedance data. As $\beta$ increases the maximum and minimum of
$R_s(T)$ change into an inflection point with a horizontal tangent
and for larger $\beta$ values the maximum of $R_s(T)$ disappears
completely \cite{Fink1}.

The linear $T$--increase of $R_s(T)$ at low temperatures  (inset
in Fig.~5) is a direct consequence of the linear change of $\lambda(T)$
near $T=0$, proportional to the coefficient $\alpha$ in Eq.~(\ref{Llin}),
and due to a constant scattering rate at low temperatures, as shown in
Fig.~4.

The dashed and dotted curves in Fig.~5, are calculated
$R_s(T)$ values at 1.14~GHz, with $t^5$ replaced
by $t^4$ (dashed curve) and by $t^6$ (dotted curve) in Eq. (\ref{Tau1}).
The best fit of the experimental data is  $1/\tau(t)\propto t^5$. Moreover,
Eq.~(\ref{Tau2}) enables us to incorporate the shoulder of $R_s(T)$ obtained
 with YBCO single crystals in Refs.~\cite{Srik1,Srik2}. This is shown in
Fig.~6, which contains measurements (squares) of $R_s(T)$ at 10~GHz taken
from Ref.~\cite{Srik1}, and calculations (solid line) of $R_s(T)$ using
Eqs.~(\ref{Sigma1}) and
(\ref{RS}) with $\omega  \tau(T_c)=4\times 10^{-3}$,
$\sigma_2(T)/\sigma_2(0)$ obtained from the same experimental data
\cite{Srik1}, $\beta=0.02$ and $\kappa=4$ in Eq.~(\ref{Tau2}) \cite{Tru2}.
\begin{figure}
\centerline{\psfig{figure=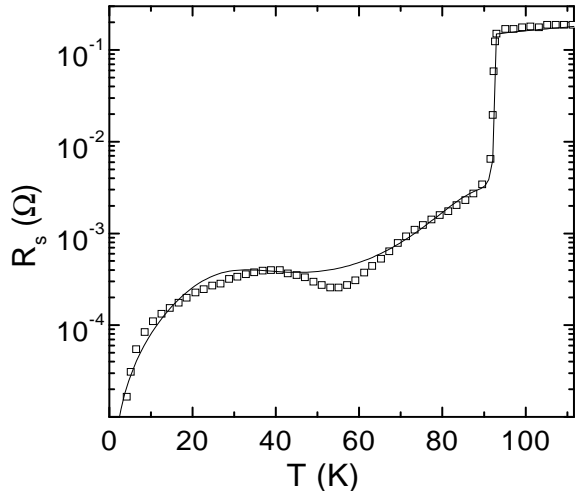,height=6.5cm,width=7.5cm,clip=,angle=0.}}
\caption{Comparison between calculated (solid line) and measured
(squares) surface resistance $R_s(T)$ of YBCO single crystal at 10~GHz.
Experimental data are from Ref.~\protect\cite{Srik1}. }
\label{r6}
\end{figure}

The calculated curves in Figs.~5 and 6 are very close to the experimental
data and display the common and unique features of $R_s(T)$ for $T<T_c$ and
$\omega\tau <1$ of high-quality YBCO single crystals fabricated
by different methods, namely:
(i) the linear temperature dependence of surface resistance,
$\Delta R_s(T)\propto T$, caused by the linear variation of
$\Delta X_s(T)\propto \Delta \lambda_{ab}(T)\propto T$ at temperatures
$T\ll T_c$, and by $\tau(T) \rightarrow {\rm const}$ at low temperatures;
(ii) the broad peak of $R_s(T)$ in the intermediate
temperature range due to the rapid decrease of the relaxation time
$\tau(T)\propto T^{-5}$, with increasing temperature; and (iii) the
increase in $R_s(T)$ in the range $T_c/2<T<T_c$ (Fig.~6) caused by the
crossover from $T^{-5}$ to $T^{-1}$  of $\tau(T)$ in Eq.~(\ref{Tau2}),
which occurs in Fig.~6 at a lower temperature than
in Fig.~5. The behavior of $1/\tau(T)$ for these two YBCO crystals is
shown in Fig.~4.

Up to this point, our analysis has not taken into account the residual
surface resistance $R_{{\rm res}}$ of the samples. In the YBCO crystals
whose data are plotted in Figs.~5 and 6, scaled to the same frequency of
10~GHz, the resistance $R_{{\rm res}}< 50$~$\mu \Omega $.
$R_{{\rm res}}/R(T_{{\rm c}})< 10^{-3}$ is so small that $R_{{\rm res}}$ was
neglected even at $T\ll T_{{\rm c}}$. In most HTS crystals which were
investigated, however, $R_{{\rm res}}/R(T_c)>10^{-3}$
(see, e.g., Figs.~1 and 2). Therefore, it is important that $R_{{\rm res}}$
is added to the calculated $R_s(T)$ values when comparing the latter with
the experimental data.

Figure~7 compares the measured $R_s(T)$ and $X_s(T)$ of BSCCO, plotted in
Fig.~1, with calculations obtained from Eqs.~(\ref{RS}) and (\ref{XS}). In
this case, we have added to the calculated values of $R_s(T)$
a constant $R_{{\rm res}}=0.5$~m$\Omega $. The calculation is based on
measurements of $\sigma _2(T)/\sigma _2(0)$ obtained in the
same experiment and plotted in the inset to Fig.~13 (see below), with
parameter $\omega \tau (T_{{\rm c}})=0.9\times 10^{-2}$, $\beta =2$ and
$\kappa =3$ in Eq.~(\ref{Tau2}). It is clear that the agreement between
the calculated and experimental curves is good
throughout the temperature interval $5\leq T\leq 120$~K.
\begin{figure}
\centerline{\psfig{figure=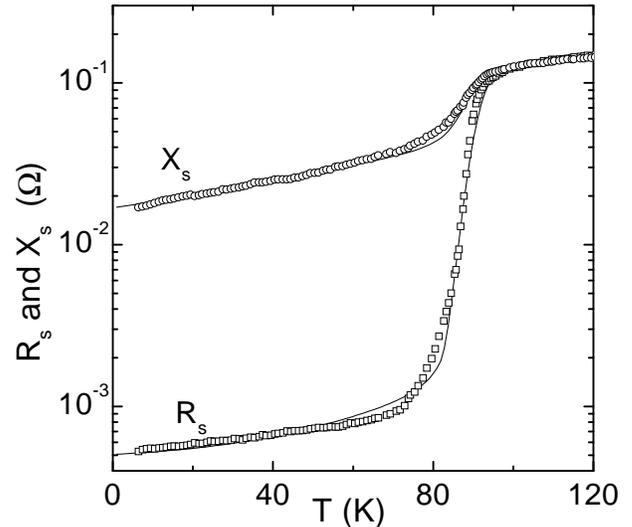,height=7cm,width=8cm,clip=,angle=0.}}
\caption{Comparison between calculated (solid lines) and measured
surface impedance (symbols) of BSCCO single crystal (see Fig.~1). A constant
$R_{{\rm res}}=0.5$~m$\Omega$ is added to the values of $R_s(T)$, obtained
from Eq.~(\protect\ref{RS}).}
\label{r7}
\end{figure}

Another reason for including $R_{{\rm res}}$ is that the ratio
$R_{{\rm res}}/R(T_{{\rm c}})\propto\omega^{3/2}$.
Fig.~8 is based on the experimental data of BSCCO
single crystal measured in Ref.~\cite{Lee} at three frequencies:
14.4~GHz ($\omega\tau(T_c) = 0.7 \times 10^{-2}$),
24.6~GHz, and 34.7~GHz. The solid curves are calculations at these
frequencies obtained from Eqs.~(\ref{Sigma1}) and (\ref{RS}) using $\tau(T)$
from Eq.~(\ref{Tau2}) with $\beta=0.1$ and $\kappa=4$. The comparison
procedure is different from that discussed above for YBCO crystals
since $R_{{\rm res}}\propto\omega^2$ is added to the calculated
$R_s(T)$ values. The inset of Fig.~8 shows a linear plot of the measured and
calculated surface resistance at low temperatures. We emphasize that at
temperatures below $T_c/2$ the value of $\Delta R_s(T)$ changes
proportional to $T$.
\begin{figure}
\centerline{\psfig{figure=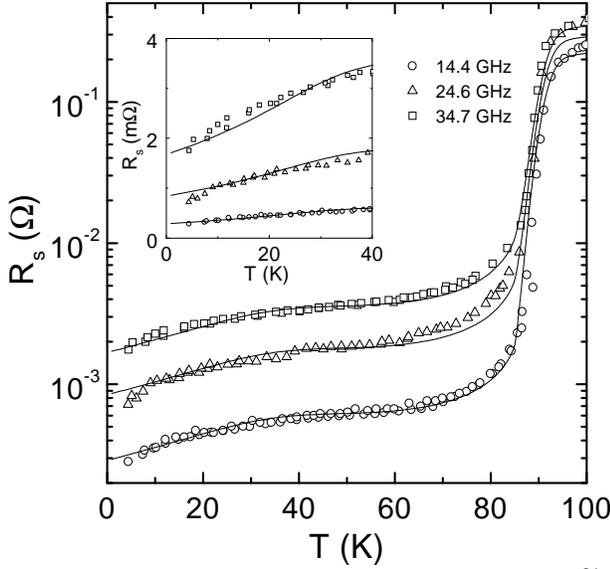,height=7.5cm,width=8cm,clip=,angle=0.}}
\caption{Experimental data of BSCCO single crystal \protect\cite{Lee}
at various frequencies: 14.4~GHz, 24.6~GHz, and 34.7~GHz. The solid curves
are the
calculated [$R_s(T) + R_{\rm{res}}$]--functions, with $R_{\rm{res}}$ values
of 0.29, 0.85 and 1.7 m$\Omega$, respectively. The inset shows the linear
temperature dependencies of the surface resistance at low temperatures.}
\label{r8}
\end{figure}

In the millimeter and shorter wavelength bands, the condition
$\omega\tau<1$ may not be satisfied in the superconducting
state of the purest HTS single crystals due to the fast growth of $\tau(T)$
with decreasing $T<T_{{\rm c}}$. Therefore, it is natural not only to
take $R_{{\rm res}}$ into account in analyzing the experimental data of
$Z_s(T)$ and $\sigma_s(T)$ but also the more general Eq.~(\ref{S1-S2}) of
the two-fluid model should replace Eq.~(\ref{Sigma1}). The $R_s(T)$ data
of Ref.~\cite{Bon100} at frequencies of 13.4, 22.7, and 75.3~GHz, are shown
in Fig. 9 with the calculated $R_s(T)$ values (obtained on the same YBCO
crystal as was used in Fig.~5). We used
$\tau(T_c)/\tau(0) \approx \beta = 5\times 10^{-3}$ in Eq.~(\ref{Tau1})
for all curves\linebreak
\begin{figure}
\centerline{\psfig{figure=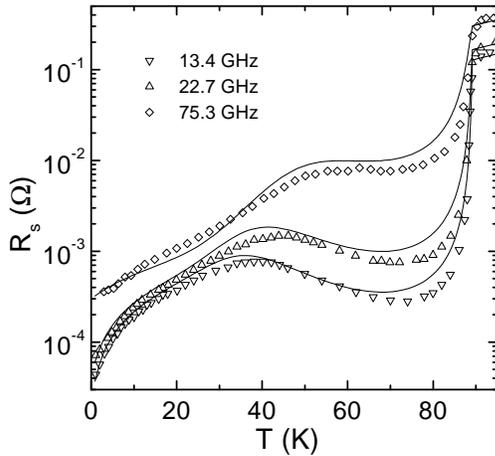,height=6cm,width=6.5cm,clip=,angle=0.}}
\caption{Comparison between calculated (lines) and measured
\protect\cite{Bon100} (symbols)  surface resistance $R_s(T)$
of YBCO single crystal at 13.4, 22.7 and 75.3~GHz. We assumed
$R_{\rm{res}}=0.3$ m$\Omega$ for $75.3$ GHz, zero for the other frequencies.}
\label{r9}
\end{figure}
\noindent
shown in Fig.~9 (same as previously used in Fig. 5), and
added $R_{{\rm res}}=0.3$~m$\Omega$  to $R_s(T)$ [Eq.~(\ref{RS})]
at  75.3~GHz only. The conductivity components $\sigma_1(T)$
and $\sigma_2(T)$ which are contained in Eq.~(\ref{RS}) are obtained
from the experimental data of $\sigma_2(T)/\sigma_2(0)$ at 1.14~GHz,
\cite{Bon100} (shown in Fig.~11), and from Eq.~(\ref{S1-S2}).

Figure 10 shows another example. The experimental  $R_s(T)$ data (squares)
of TBCO single crystal ($T_c=78.5$~K) \cite{Broun} are compared with
calculations based on Eqs.~(\ref{RS}), (\ref{S1-S2}), and (\ref{Tau2}).
The curve representing the theoretical $R_s(T)+R_{{\rm res}}$ was plotted
using $\beta=0.1$, $\kappa =5.5$,
$\omega \tau (T_{{\rm c}})=1.7\times 10^{-2}$,
$R_{{\rm res}}=0.8$~m$\Omega$, and with $\sigma_2(T)/\sigma_2(0)$, shown
in the inset (circles) of Fig.~10.

\begin{figure}
\centerline{\psfig{figure=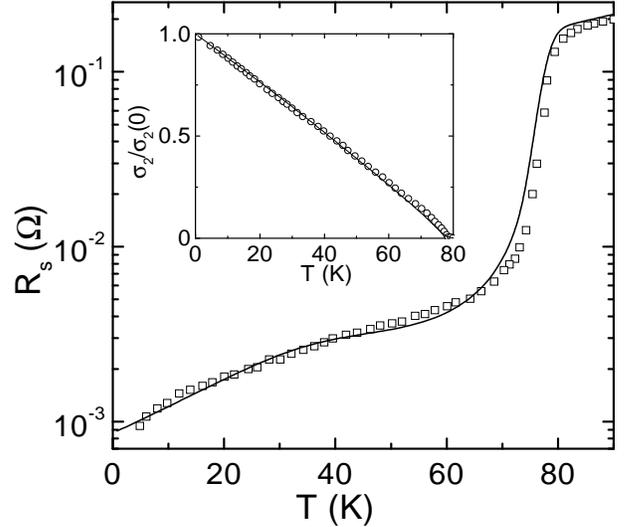,height=7cm,width=8cm,clip=,angle=0.}}
\caption{Surface resistance $R_s(T)$ of a TBCO single crystal
at 24.8~GHz taken from Ref.~\protect\cite{Broun}
Solid curve is the calculated [$R_s(T)+R_{\rm{res}}$]--function with
$R_{\rm{res}}=0.8$~m$\Omega$.
The inset shows  measured \protect\cite{Broun} (circles)
 and calculated results of $\sigma_2(T)/\sigma_2(0)$ (solid line),
using Eq.~(\protect\ref{ns1}) with $\alpha=0.9$. }
\label{r10}
\end{figure}

\subsection{Temperature dependence of the superconducting electron density}

Our phenomenological model would be incomplete if simple
formulas were not available that describe correctly the measurements of
$\Delta\lambda_{ab}(T)$.
Figures~10 (inset), 11 and 12 show
$\sigma_2(T)/\sigma_2(0)=\lambda^2(0)/\lambda^2(T)=n_s(T)/n$
in the $ab$-plane of TBCO, YBCO, and BSCCO single crystals from
Refs.~\cite{Broun,Bon100} and \cite{Lee}, respectively.
All of these quantities change linearly with temperature at low-temperatures
and can be approximated by the function \cite{Tru3}

\begin{equation}
n_s/n=(1-t)^\alpha,  \label{ns1}
\end{equation}

\noindent
where $\alpha$ is a numerical parameter. For $t\ll 1$, Eq.~(\ref{Llin})
follows from Eq.~(\ref{ns1}). For the cited  experiments, the values of
$\alpha$ fall into the range
 $0.4< \alpha \le 0.9$. Near $T_c$ we obtain
$\lambda(t)\propto n_s(t)^{-1/2}\propto (1-t)^{-\alpha/2}$, which is
also in fairly good agreement with experimental data. However,
equation~(\ref{ns1}) yields an infinite value of derivative
$d\sigma_2(t)/dt\propto (1-t)^{\alpha-1}$ at $t=1$ for $\alpha<1$.
\begin{figure}
\centerline{\psfig{figure=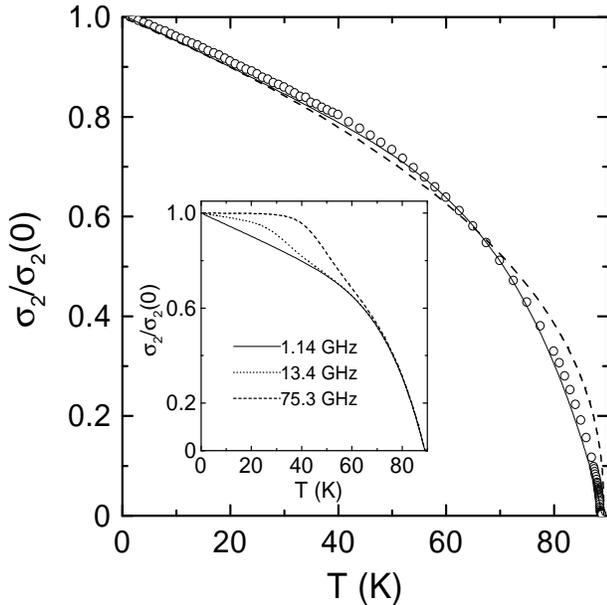,height=8cm,width=8cm,clip=,angle=0.}}
\caption{Plots of Eq.~(\protect\ref{ns1}) (dashed line, $\alpha=0.42$)
and Eq.~(\protect\ref{ns2}) (solid line, $\alpha=0.47$), showing the fit
to the empirical $\sigma_2(T)/\sigma_2(0)$. The
experimental data (circles) are from Ref.~\protect\cite{Bon100} at 1.14~GHz.
The inset shows the temperature dependencies of $\sigma_2(T)/\sigma_2(0)$
at various frequencies, calculated from
 Eqs.~(\protect\ref{S1-S2}), (\protect\ref{ns2}) and
(\protect\ref{Tau1}) . }
\label{r11}
\end{figure}

\begin{figure}
\centerline{\psfig{figure=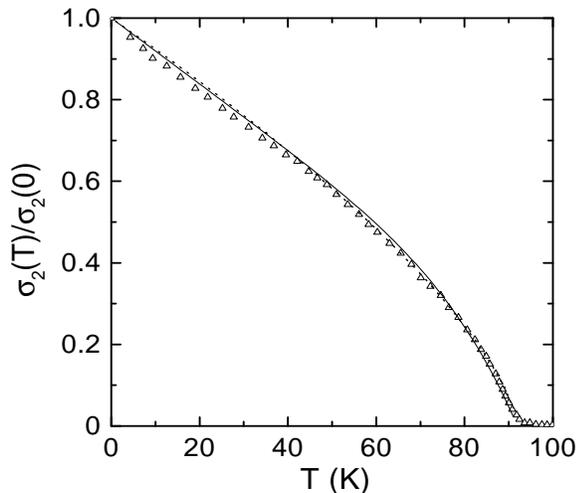,height=6.5cm,width=7.5cm,clip=,angle=0.}}
\caption{Comparison between calculated (solid curve:
Eq.~(\protect\ref{ns2}), $\alpha=0.74$; and, dotted line:
Eq.~(\protect\ref{ns1}), $\alpha=0.7$) and measured \protect\cite{Lee}
(symbols) of $\sigma_2(T)/\sigma_2(0)$ values of  BSCCO
single crystal \protect\cite{Fink2}. }
\label{r12}
\end{figure}

An approximation for $n_s(t)/n$ close to Eq.~(\ref{ns1})
was proposed in Ref.~\cite{Fink2}:

\begin{equation}
n_s/n=1-\alpha t - (1-\alpha)\,t^6  \label{ns2}
\end{equation}

\noindent
and is shown by solid lines in Figs.~11 and 12. Equation (\ref{ns2}) insures
that the slope at $T_c$ of $\lambda^2(0)/\lambda^2(t)|_{T_c}=(5\alpha-6)$
is finite and negative for $\alpha<1.2$.

The above functions for $n_s(t)$, however, in their simplest forms
(\ref{ns1}) and (\ref{ns2}), cannot account for all features in
$\lambda^2(0)/\lambda^2(T)$ detected recently in YBCO crystals (see Table~I)
in the intermediate temperature range \cite{Tru2,Srik1,Tru4,Srik2}.
Moreover, the slope of these curves at $T\ll T_c$ requires that $\alpha >1$
in Eq.~(\ref{ns1}), which
would lead to  zero slope of the $\sigma_2(T)/\sigma_2(0)$ curve at $T=T_c$.
Therefore  we have added an additional empirical term to the right-hand side
of  Eq.~(\ref{ns1}) without violating the condition of particle conservation,
$n_s+n_n=n$,

\begin{equation}
n_s/n=(1-t)^{\alpha}(1-\delta)+\delta(1-t^{4/\delta}),  \label{ns3}
\end{equation}

\noindent
where $0< \delta < 1$ is the weight factor \cite{Tru2}. For $\delta\ll 1$
and $\alpha >1$ the dominant contribution to $\sigma_2(T)$ throughout the
relevant temperature range is still due to the first term on the right of
Eq.~(\ref{ns3}), while the second is responsible  for the  finite slope
of $\sigma_2(T)/\sigma_2(0)$ at $T=T_c$, equal to $-4$, in accordance with
the GC model. As $\delta$ increases, the second term on the right side
of Eq.~(\ref{ns3}) becomes more essential. The experimental curve of
$\sigma_2(T)/\sigma_2(0)$, derived from $R_s(T)$ and $X_s(T)$
measurements of YBCO crystal in Ref.~\cite{Tru2}, is properly described
by Eq. (\ref{ns3}) with $\delta=0.5$ and $\alpha=5.5$ (Fig.~13). This
calculation reflects the characteristic features of the experimental data,
namely, the linear section of $n_s$ and the positive second derivative
($\alpha>1$) in
the low-temperature region, the plateau in the intermediate temperature
range, and the correct value of the slope near $T_c$.

Using Eq. (\ref{ns3}) with $\alpha=2$ and $\delta=0.2$, one can also
describe the $T$--dependence of $\sigma_2(T)$ of
BSCCO crystals (Figs.~1 and 7), plotted in the inset to Fig.~13.
\begin{figure}
\centerline{\psfig{figure=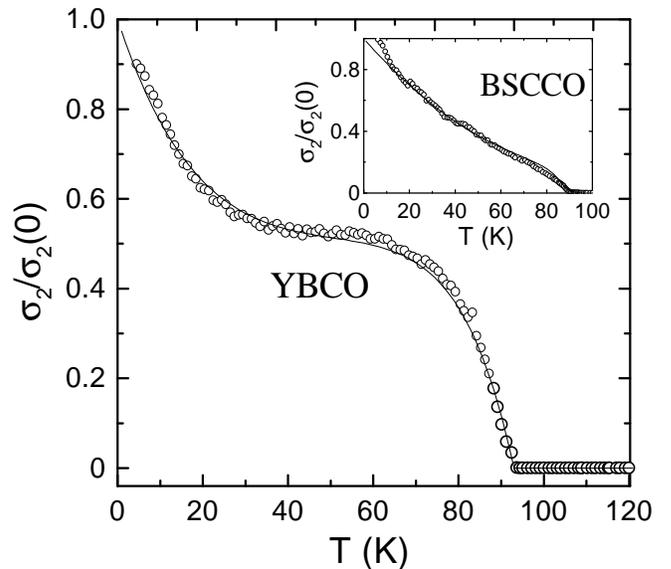,height=7.5cm,width=8.5cm,clip=,angle=0.}}
\caption{Comparison between calculated (solid line) and measured
(circles) values of $\sigma_2(T)/\sigma_2(0)$ of YBCO single crystal
\protect\cite{Tru2}. The inset shows the measured and calculated values, using
Eq.~(\protect\ref{ns3}) for the temperature
 dependencies of $\sigma_2(T)/\sigma_2(0)$
of the BSCCO crystal, shown in Fig.~1. }
\label{r13}
\end{figure}

\subsection{Real part of conductivity}

Since the measurements and calculations
of $R_s(T)$, $X_s(T)$, and $\sigma_2(T)$ are in good agreement and consistent
with  $\sigma_1(T)$ in the range $T<T_c$, it is proposed that
the modified two-fluid model is a powerful tool for describing the
electrodynamic properties of HTS's. The only feature that has not been
investigated by this model is the behavior of $Z_s(T)$ and $\sigma_s(T)$
in the temperature range near $T_{{\rm c}}$. A spectacular display is the
narrow peak in the real part of the conductivity (see Fig.~3).

$\sigma _1(T)$ of YBCO crystals, obtained from measurements at
1.14~GHz \cite{Bon100}, is plotted (circles) in Figs.~3 and 14.
\begin{figure}
\centerline{\psfig{figure=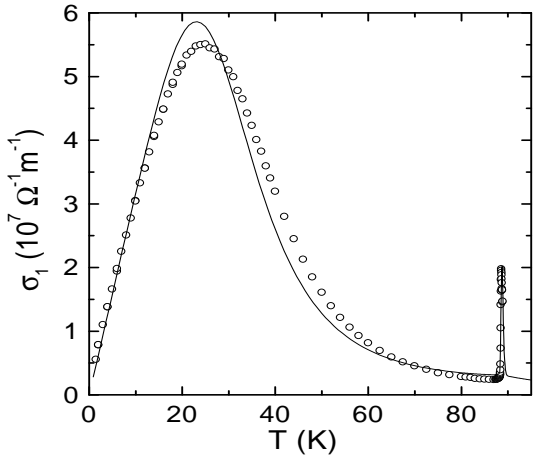,height=6cm,width=7cm,clip=,angle=0.}}
\caption{Comparison of the experimental $T$--dependence of $\sigma_1(T)$ (open
circles in Fig.~3) of YBCO single crystal at 1.14~GHz
(Ref.~\protect\cite{Bon100}) with that calculated using the modified
two-fluid model (solid line), taking into account the inhomogeneous
broadening of the superconducting transition ($\delta T_c=0.4$~K in
Eq.~(\protect\ref{zef})). }
\label{r14}

The narrow peak near $T_c$ can be described by an  effective
medium model \cite{Gol1,Tru1} which takes into account
inhomogeneous broadening of the superconducting transition.
Assume that different regions of a given specimen experience transitions
to the superconducting state at different temperatures within the $T$--range
$\delta T_c$. If the dimension of each of these regions is smaller than the
magnetic field penetration depth (microscopic-scale disorder), the
distribution of the microwave currents over the sample is uniform, and the
calculation of the effective impedance $Z_{{\rm eff}}$ of the sample reduces
to two operations: first, the impedances $Z_s$ of all regions in the
specimen (with different $T_c$) that are connected in series along a current
path are added, and, second, averaging over the sample volume is performed.
As a result, we obtain

\begin{equation}
Z_s^{{\rm eff}}(T)=R_s^{{\rm eff}}(T)+iX_s^{{\rm eff}}(T)
=\int\limits_{\delta T_c}{Z_s}(T,T_c)f(T_c)dT_c\ ,
\label{zef}
\end{equation}

\noindent
where the distribution function $f(T_c)$ is such that the fraction of the
sample volume with critical temperatures in the range $T_c<T<T_c+dT_c$
equals $f(T_c)dT_c$. In the simplest case $f(T_c)$ is a Gaussian function.
In the experiments of Ref. \cite{Bon100}, the width of the superconducting
resistive transition was approximately 0.4~K, which we equate to the
width of the Gaussian distribution $f(T_{{\rm c}})$. Using the general
relations (\ref{S-R}), with the effective impedance components obtained
from Eq.~(\ref{zef}), $\sigma _1^{{\rm eff}}(T)$ is calculated near $T_c$
and is plotted with the experimental data in Fig.~14 for YBCO \cite{Bon100}.
The overall agreement is good.
\end{figure}

In the framework of the discussed approach, $\sigma _1^{{\rm eff}}(T)$
displays a narrow peak at $T^*=T_c-\delta T_c$. It is easy to check
that the relative peak amplitude is approximately equal to

\begin{equation}
\frac{\sigma_1(T^*)-\sigma (T_c)}{\sigma (T_c)}\approx
\left\{\begin{array}{l}
\gamma,\,\, \mbox{ if } \gamma>1\\
\gamma^2, \mbox{ if } \gamma<0.1
\end{array}\right.\;,
\label{sef}
\end{equation}

\noindent
where $\gamma = \delta T_c/[T_c\,\omega\tau(T_c)]$, implying, the narrower
the superconducting resistive transition, the smaller the peak amplitude.
Usually, experiments yield $\gamma>1$ (e.g., the data of Ref.~\cite{Bon100}
gives $\gamma\simeq 7$ at 1.14~GHz) and, therefore, the peak
amplitude should be inversely proportional to frequency.

We applied the above procedure to other specimens to incorporate corrections
into the calculations of the $\sigma_1(T)$ curves, caused by inhomogeneous
broadening of the superconducting transition.
We adjusted the previous calculations of $R_s(T)$ (Figs.~7, 8, and 10)
and $\sigma_2(T)$ (Figs.~10, 12, and inset to Fig.~13) by substituting
the resulting $Z_s^{{\rm eff}}(T)$ into the general equation~(\ref{S-R})
for the conductivity $\sigma_1$. The resulting curves for BSCCO and TBCO
 are shown in Figs.~15--17.
\begin{figure}
\centerline{\psfig{figure=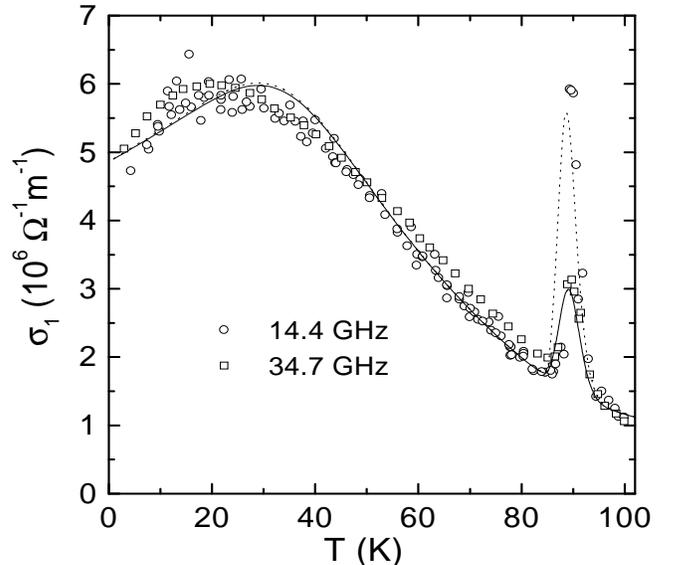,height=7.5cm,width=8.5cm,clip=,angle=0.}}
\caption{Experimental data of $\sigma_1(T)$ at 14.4 and 34.7~GHz
of BSCCO single crystal
\protect\cite{Lee} and calculations of $\sigma_1(T)$ using
Eqs.~(\protect\ref{Tau1}), (\protect\ref{RS}), (\protect\ref{zef})
and (\protect\ref{S1-S2}), taking into account sample inhomogeneities
($\delta T_c=2$~K). }
\label{r15}
\end{figure}

\begin{figure}
\centerline{\psfig{figure=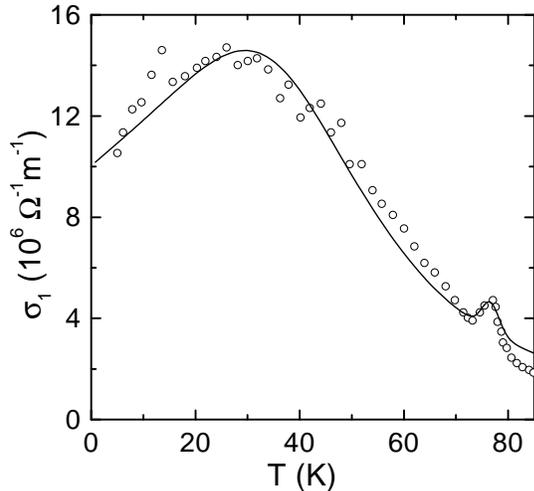,height=6.5cm,width=7cm,clip=,angle=0.}}
\caption{Comparison of the $T$--dependence of the
experimental $\sigma_1(T)$ (open
circles) of TBCO single crystal at 24.8~GHz (Ref.~\protect\cite{Broun})
with that calculated using the modified
two-fluid model (solid line), taking into account the inhomogeneous
broadening of the superconducting transition ($\delta T_c=2.5$~K in
Eq.~(\protect\ref{zef})). }
\label{r16}
\end{figure}

\begin{figure}
\centerline{\psfig{figure=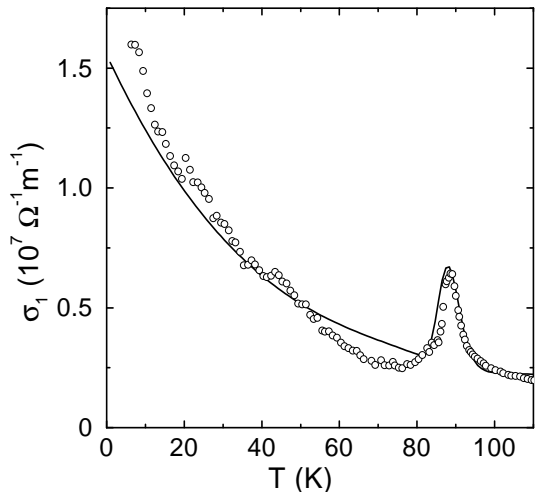,height=6.5cm,width=7cm,clip=,angle=0.}}
\caption{Conductivity $\sigma_1(T)$ of BSCCO single crystal at $9.4$ GHz,
extracted from the surface impedance measurements of Fig.~1, and calculation
based on the modified two-fluid model, which takes into account the
inhomogeneous broadening of the superconducting transition
($\delta T_c=4.5$~K). $\sigma_1(T)$ does not have
a broad peak at low temperatures in this particular case.}
\label{r17}
\end{figure}

\section{Conclusion}

We have presented a summary of measurements of the surface
impedance  $Z_s(T)=R_s(T)+iX_s(T)$ of high-quality
YBCO, BSCCO, TBCO and TBCCO crystals in  the superconducting and normal
states (Table~I). For frequencies $\alt 10$~GHz the common features of all
these materials are the linear temperature dependence of surface resistance,
$\Delta R_s(T)\propto T$, of the surface reactance,
$\Delta X_s(T)\propto \Delta \lambda_{ab}(T)\propto T$, at temperatures
$T\ll T_c$, their rapid growth as $T\to T_c$, and their behavior in
the normal state corresponding to a linear $T$--dependence of
$\Delta\rho _{ab}(T)$, with $R_s(T)=X_s(T)=\sqrt{\omega \mu_0\rho (T)/2}$.
There are differences between the $T$--dependence
of $Z_s(T)$ of BSCCO, TBCCO or TBCO single crystals, with
tetragonal lattices, compared to YBCO crystals, with
an orthorhombic lattice. The linear resistivity region
of the tetragonal materials extends to near $T_c/2$, while for YBCO the
linear region terminates near or below $T<T_c/3$.
At higher temperature,  $R_s(T)$ of YBCO has a broad peak. In addition,
the $\lambda _{ab}(T)$ curves of some YBCO single crystals have unusual
features in the intermediate temperature range.

We describe all of the above features of $Z_s(T)$ and
$\sigma_s(T)=\sigma_1(T)-i\sigma_2(T)=i\omega\mu_o/Z_s^2(T)$
of high-quality HTS crystals by generalizing the well-known GC
two-fluid model:

\noindent
(i) we introduce a temperature dependence of the relaxation
time of the quasiparticles in accordance with the Bloch-Gr\"uneisen
law. We find that the $R_s(T)$ curves in different HTS crystals
are well described using Eqs.~(\ref{Tau1}) or (\ref{Tau2}) for
$1/\tau(T)$. In the latter equation there is only one fitting parameter,
$\kappa=\Theta /T_c$, while the other parameter
$\beta =\tau(T_c)/\tau(0)\ll 1$ can be estimated directly from
the experimental data with the help of Eqs.~(\ref{Omtc}) and (\ref{Omt0}).
The absence of the broad peak of $R_s(T)$ in tetragonal HTS single
crystals is due to a less rapid increase of $\tau(T)$ with decreasing
temperature. In other words, the value of $\beta$ is smaller for
YBCO crystals than for BSCCO, TBCO or TBCCO. For the latter crystals
the residual losses $R_{\rm{res}}$ are usually large and they
have to be taken into account.

\noindent
(ii) we replace the well-known temperature dependence of the density
of superconducting carriers in the GC model, $n_s=n(1-t^4)$, by one of
the functions proposed by Eqs. (\ref{ns1}), (\ref{ns2}) or (\ref{ns3}).
All of these functions change linearly with temperature at $t\ll 1$
(see Eq.~(\ref{Llin})). This permits one to extract the common and
distinctive features of $X_s(T)$ and $\sigma_2(T)$ from different HTS
crystals.

It also follows from the equations of the modified two-fluid model, that
at low temperatures, $t\ll 1$, and low frequencies ($\omega\tau(0) < 1$),
all curves of $Z_s(T)$ and $\sigma_s(T)$ have linear
regions: $\sigma_1\propto \alpha t/\beta$, since $n_n/n\approx \alpha t$
and $\tau\approx \tau(0)\approx \tau(T_c)/\beta$. Furthermore,
$\Delta \sigma_2\propto -\alpha t$. Then, in accordance with Eq.~(\ref{R-X}),
$R_s\propto \alpha t/\beta$ and
$\Delta X_s\propto \Delta \lambda \propto \alpha t/2$.
As the temperature increases, the curve of $\sigma_1(t)$ passes through a
maximum at $t \alt 0.5$ if the unequality (\ref{Rres}) is valid. This peak is
due to superposition of two competing effects, namely, the decrease in the
number of normal carriers as the temperature decreases, for $t<1$, and the
increase in the relaxation time, which saturates at $t\sim\beta^{1/5}$,
at which point the impurity scattering starts to dominate.
The features in the $X_s(T)$ and $\sigma_2(T)$ curves for YBCO single
crystals in the intermediate temperature range (plateau \cite{Tru2}
or bump \cite{Srik1}) can also be described within the framework of our
modified two-fluid model, if we take into account the modification
 of $n_s(t)$, described by Eq.~(\ref{ns3}) with $0<\delta\le 0.5$.
The narrow peak in the real part of the conductivity
$\sigma_1(T)$, near $T_c$, in HTS single crystals can be explained
in terms of an effective medium model, taking into account
strong electron-phonon coupling of the quasiparticles and
inhomogeneous broadening of the superconducting transition.

It is natural to compare the tenets of our phenomenological model with
the results of microscopic theories. As was shown in
Refs.~\cite{Tru9} and \cite{Tru100}, the simple formula (\ref{Llin}),
which defines the linear low temperature dependence
of the magnetic field penetration depth in the $ab$-plane of HTS crystals,
is consistent with the $d$-wave model \cite{Pin1,Pin2,Pin3}
in the limit of strong (unitary) scattering \cite{Hir2}.
Besides, there is nothing foreign in introducing the function
$1/\tau(T)\propto T^5$ for the purpose of characterizing scattering
in the superconducting state of HTS. A similar temperature dependence of
the relaxation rate of quasi-particles follows from the SC model if the
phonon corrections to the electromagnetic vertex are taken into account
\cite{El4}.

In the framework of our modified two-fluid model, the linear low
$T$--dependence of
the real part of conductivity $\sigma_1(T)$ is consistent with a constant
scattering rate, as it is in a normal metal. While the assumption of a Drude
form of the conductivity is supported by the $d$-wave microscopic
analysis \cite{Hir2}, it was shown that
pair correlations in the usual impurity scattering models lead to strong
$T$--dependence of the scattering time (neglecting vertex corrections),
namely, $\tau (T)\propto T$ in unitary limit, or $\tau (T)\propto 1/T$ in
the Born limit. An attempt to resolve this problem in Ref.\cite{Hens} by
choosing an intermediate scattering rate has not provided satisfactory results
yet. Very recently the authors of Refs.~\cite{Hettler} and \cite{Berlin}
argue that experimental observation $\sigma_1(T)\propto T$ could be explained
by the generalized Drude formula $\sigma_1(T)\propto n_{qp}(T)\tau(T)$ if
the quasiparticle density varies as $n_{qp}(T)\propto T$ (as indeed happens
for $d$-wave pairing) and if the effective quasiparticle scattering
time $\tau(T)$ saturates at low $T$. Various possible physical
mechanisms of the temperature and energy dependence of $\tau $ are
discussed in \cite{Hettler,Berlin}: scattering from the ''holes'' of the
order parameter at impurity sites, and scattering from extended defects.
These mechanisms may provide the required saturation of $\tau (T)$ at
low $T$. As was discussed recently in Ref.~\cite{DurLee}, the vertex
corrections can also modify the low temperature conductivity. However,
the temperature dependence has not been investigated yet.

Nevertheless, the microscopic models, which have been investigating
the microwave response based on
a pure $d$-wave order parameter symmetry, cannot account for the linear
section of the $R_s(T)$ curves extending to $T_c/2$ (at frequencies of
10~GHz and below) in tetragonal HTS
single crystals, observation of radically different values of the
slopes of $\sigma _2(T)$ for $T\ll T_c$ (corresponding to $\alpha >1$ in
Eq.~(\ref{ns3})), observed on YBCO crystals
\cite{Tru2,Srik1,Tru4,Srik2,Tru5}, and unusual features of $\sigma_2(T)$
in the intermediate temperatures range.

Recently, observations of unusual microwave properties of HTS materials
have caught the attention of a number of
researchers \cite{Comb,Don1,Kim,Zaz1,Pok,Modr,YuraLT22}.
These observations are tentatively attributed to mixed
$(d+s)$-wave symmetry of the order parameter. Most studies deal with
 the low temperature variation of the London penetration depth  and
its relation to an order parameter of
 mixed symmetry. In particular,
it was shown in Ref.~\cite{Modr} that the low temperature properties of
$\lambda (T)$ can be used to distinguish between a pure $d$-wave order
parameter and one with $(s+id)$ symmetry, having a small subdominant $s$-wave
contribution in systems connected with a tetragonal lattice. Moreover,
additions of impurities suppress the $d$-wave symmetric part to the benefit
of the $s$-wave part. As a result, a variety of low-temperature dependencies
of $\lambda(T)$ is
possible for various impurity concentrations, which allows one, in principle,
to determine whether or not the order parameter of a superconductor with an
orthorhombic lattice is of ($s+id$)  or ($s+d$) symmetry
\cite{Beal1}. In Ref.\cite{Zaz1} the ($d+s$) model was generalized
 to take into account the normal state anisotropy. This is the
realistic approach to high-$T_c$ cuprates with an orthorhombic
distortion, since recent microwave conductivity data suggest that a
substantial part of the $ab$-anisotropy of $\lambda (T)$ is a normal
state effect. It was shown that such an anisotropy affects not only
the $ab$-anisotropy of the transport coefficients, but also the density
of states and other thermodynamic quantities. The possible temperature
variation of the penetration depth $\lambda (T)$ was analyzed recently
in Ref.~\cite{YuraLT22} in the framework of the ($d+s$) model of hybrid
pairing. The slope of the $\Delta \lambda (T)\propto T$ for $T\ll T_c$
and its dependence on the $\Delta _s/\Delta _d$ admixture in the gap function
was analyzed quantitatively, taking into account the impurity scattering.
However, the quantitative comparison of the latter calculation with the
experimental data has not been performed yet. More interesting discoveries
in this field of research can be expected in the near future.

\section{Acknowledgments}

We are greatly indebted to N.~Bontemps, V.~T.~Dolgopolov, V.~F.~Gantmakher,
A.~A.~Golubov, L.~M.~Fisher, E.~G.~Maksimov, and V.~P.~Mineev for many
helpful discussions.
H.~J.~F. thanks D.~A.~Bonn for permission to use the original
data points of Ref. \cite{Bon100} in our figures. The research of Yu.~A.~N.
and M.~R.~T. has been supported by by the Russian Fund for Basic Research
(grant 97-02-16836) and Scientific Council on Superconductivity (project
96060), and in part by the Program for Russian-Dutch Research Cooperation
(NWO).

\end{multicols}


\begin{references}

\bibitem{Har}  W.~N.~Hardy, D.~A.~Bonn, D.~C.~Morgan, R.~Liang, and
K.~Zhang, Phys. Rev. Lett. {\bf 70}, 3999 (1993).

\bibitem{Ach}  D.~Achir, M.~Poirier, D.~A.~Bonn, R.~Liang, and W.~N.~Hardy,
Phys. Rev. B {\bf 48}, 13184 (1993).

\bibitem{Kam}  S.~Kamal, D.~A.~Bonn, N.~Goldenfeld, P.~J.~Hirschfeld,
R.~Liang, and W.~N.~Hardy, Phys. Rev. Lett. {\bf 73}, 1845 (1994).

\bibitem{Bon3}  D.~A.~Bonn, S.~Kamal, K.~Zhang, R.~Liang, D.~J.~Baar,
E.~Klein, and W.~N.~Hardy, Phys. Rev. B {\bf 50}, 4051 (1994).

\bibitem{Mao}  J.~Mao, D.~H.~Wu, J.~L.~Peng, R.~L.~Greene, and S.~M.~Anlage,
Phys. Rev. B {\bf 51}, 3316 (1995).

\bibitem{Bon33}  D.~A.~Bonn, S.~Kamal, K.~Zhang, R.~Liang, and W.~N.~Hardy,
J. Phys. Chem. Solids {\bf 56}, 1941 (1995).

\bibitem{Jac}  T.~Jacobs, S.~Sridhar, C.~T.~Rieck, K.~Scharnberg, T.~Wolf,
and J.~Halbritter, J. Phys. Chem. Solids {\bf 56}, 1945 (1995).

\bibitem{Tru2}  M.~R.~Trunin, A.~A.~Zhukov, G.~A.~Emel'chenko, and
I.~G.~Naumenko, {\it Pis'ma Zh.~Exp.~Teor.~Fiz.} {\bf 65}, 893 (1997) [JETP
Lett. {\bf 65}, 938 (1997)].

\bibitem{Srik1}  H.~Srikanth, B.~A.~Willemsen, T.~Jacobs, S.~Sridhar,
A.~Erb, E.~Walker, and R.~Fl\"{u}kiger, Phys. Rev. B {\bf 55}, R14733 (1997).

\bibitem{Tru4}  A.~A.~Zhukov, M.~R.~Trunin, A.~T.~Sokolov, and
N.~N.~Kolesnikov, {\it Zh.~Exp.~Teor.~Fiz.} {\bf 112}, 2210 (1997) [JETP
{\bf 85}, 1211 (1997)].

\bibitem{Srik2}  H.~Srikanth, Z.~Zhai, S.~Sridhar, A.~Erb, and E.~Walker,
Phys. Rev. B {\bf 57}, 7986 (1998).

\bibitem{Tru5}  M.~R.~Trunin, A.~A.~Zhukov, and A.~T.~Sokolov, J. Phys.
Chem. Solids {\bf 59}, 2125 (1998).

\bibitem{Kam8}  S.~Kamal, R.~Liang, A.~Hosseini, D.~A.~Bonn, and
W.~N.~Hardy, Phys. Rev. B {\bf 58}, 8933 (1998).

\bibitem{Bon100}  A.~Hosseini, R.~Harris, S.~Kamal, P.~Dosanjh, J.~Preston,
R.~Liang, W.~N.~Hardy, and D.~A.~Bonn, Phys. Rev. B {\bf 60}, 1349 (1999).

\bibitem{Vau}  L.~A.~de~Vaulchier, J.~P.~Vieren, Y.~Guldner, N.~Bontemps,
R.~Combescot, Y.~Lemaitre, and J.~C.~Mage, Europhys. Lett. {\bf 33}, 153
(1996).

\bibitem{Hens} S.~Hensen, G.~M\"{u}ller, C.~T.~Rieck, and K.~Scharnberg,
Phys. Rev. B {\bf 56}, 6237 (1997).

\bibitem{Vau1}  S.~Djordjevich, L.~A.~de~Vaulchier, N.~Bontemps,
J.~P.~Vieren, Y.~Guldner, S.~Moffat, J.~Preston, X.~Castel,
M.~Guilloux-Viry, and A.~Perrin, Eur. Phys. J. B {\bf 5}, 847 (1998).

\bibitem{Efim} E. Farber, G.~Deutscher, J.~P.~Contour, and E.~Jerby,
Eur. Phys. J. B {\bf 7}, (1999).

\bibitem{Jac1}  T.~Jacobs, S.~Sridhar, Q.~Li, G.~D.~Gu, and N.~Koshizuka,
Phys. Rev. Lett. {\bf 75}, 4516 (1995).

\bibitem{Shib}  T.~Shibauchi, N.~Katase, T.~Tamegai, and K.~Uchinokura,
Physica C {\bf 264}, 227 (1996).

\bibitem{Lee}  S-F.~Lee, D.~C.~Morgan, R.~J.~Ormeno, D.~M.~Broun,
R.~A.~Doyle, and J.~R.~Waldram, Phys. Rev. Lett. {\bf 77}, 735 (1996).

\bibitem{Tru90}  D.~V.~Shovkun, M.~R.~Trunin, A.~A.~Zhukov, N.~Bontemps,
H.~Enriquez, A.~Buzdin, and T.~Tamegai, to be published.

\bibitem{Broun}  D.~M.~Broun, D.~C.~Morgan, R.~J.~Ormeno, S.~F.~Lee,
A.~W.~Tyler, A.~P.~Mackenzie, and J.~R.~Waldram, Phys. Rev. B {\bf 56},
R11443 (1997).

\bibitem{Waldr}  J.~R.~Waldram, D.~M.~Broun, D.~C.~Morgan, R.~Ormeno, and
A.~Porch, Phys. Rev. B {\bf 59}, 1528 (1999).

\bibitem{Pin1}  A.~Millis, H.~Monien, and D.~Pines, Phys. Rev. B {\bf 42},
167 (1990).

\bibitem{Pin2}  H.~Monien, P.~Monthoux, and D.~Pines, Phys. Rev. B {\bf 43},
275 (1991).

\bibitem{Pin3}  P.~Monthoux, A.~Balatsky, and D.~Pines, Phys. Rev. B
{\bf 46}, 14803 (1992).

\bibitem{Hir1}  P.~J.~Hirschfeld and N.~Goldenfeld, Phys. Rev. B {\bf 48},
4219 (1993).

\bibitem{Car1}  J. P.~Carbotte and C.~Jiang, Phys. Rev. B {\bf 48}, 4231
(1993).

\bibitem{Won}  H.~Won and K.~Maki, Phys. Rev. B {\bf 49}, 1397 (1994).

\bibitem{Hir2}  P.~J.~Hirschfeld, W.~O.~Putikka, and D.~J.~Scalapino, Phys.
Rev. Lett. {\bf 71}, 3705 (1993); Phys. Rev. B {\bf 50}, 4051 (1994).

\bibitem{Scal}  D.~J.~Scalapino, Phys. Rep. {\bf 250}, 329 (1995).

\bibitem{Legg1}  J.~Annett, N.~Goldenfeld, and A.~Leggett, in Physical
Properties of High Temperature Superconductors V, D.M.~Ginsberg, eds. (World
Scientific, Singapore, 1996).

\bibitem{Maki}  K.~Maki and H.~Won, J. Phys. I France {\bf 6}, 2317 (1996).

\bibitem{Izu}  Yu. A. Izyumov, {\it Uspekhi Fiz.~Nauk}, {\bf 167}, 465
(1997); {\bf 169}, 225 (1999).

\bibitem{Klem}  R.~A.~Klemm and S.~H.~Liu, Phys. Rev. Lett. {\bf 74}, 2343
(1995).

\bibitem{Kre}  V.~Z.~Kresin and S.~A.~Wolf, Phys. Rev. B {\bf 41}, 4278
(1990); {\bf 46}, 6458 (1992); {\bf 51}, 1229 (1995).

\bibitem{Zhu1}  A.~A.~Golubov, M.~R.~Trunin, A.~A.~Zhukov, O.~V.~Dolgov, and
S.~V.~Shulga, {\it Pis'ma Zh.~Exp.~Teor.~Fiz.} {\bf 62}, 477 (1995) [JETP
Lett. {\bf 62}, 496 (1995)].

\bibitem{Adr}  S.~D.~Adrian, M.~E.~Reeves, S.~A.~Wolf, and V.~Z.~Kresin,
Phys. Rev. B {\bf 51}, 6800 (1995).

\bibitem{Tru9}  M.~R.~Trunin, {\it Uspekhi Fiz.~Nauk}, {\bf 168}, 931
(1998), [Physics--Uspekhi, {\bf 41}, 843 (1998)].

\bibitem{Nkl}  N.~Klein, N.~Tellmann, H.~Schulz, K.~Urban, S.~A.~Wolf, and
V.~Z.~Kresin, Phys. Rev. Lett. {\bf 71}, 3355 (1993).

\bibitem{Joyn}  Q.~P.~Li, E.~C.~Koltenbah, and R.~Joynt, Phys. Rev. B
{\bf 48}, 437 (1993).

\bibitem{Comb}  R.~Combescot and X.~Leyronas, Phys. Rev. Lett. {\bf 75},
3732 (1995).

\bibitem{Don1}  C.~O'~Donovan and J.~P.~Carbotte, Phys. Rev. B {\bf 52},
4568 (1995); {\bf 55}, 8520 (1997).

\bibitem{Kim}  H.~Kim and E.~J.~Nicol, Phys. Rev. B {\bf 52}, 13576 (1995).

\bibitem{Zaz1}  M. T. Beal-Monod and K. Maki, Phys. Rev. B {\bf 52}, 13576
(1995); {\bf 53}, 5775 (1996); {\bf 55}, 1194 (1997).

\bibitem{Pok}  S.~V.~Pokrovsky and V.~L.~Pokrovsky, Phys. Rev. B {\bf 54},
13275 (1996).

\bibitem{Chub}  K.~A.~Musaelian, J.~Betouras, A.~V.~Chubukov, and R.~Joynt,
Phys. Rev. B {\bf 53}, 3598 (1996).

\bibitem{Ren}  Y.~Ren, J.~Xu, and C.~S.~Ting, Phys. Rev. B {\bf 53}, 2249
(1996).

\bibitem{Shap}  A. A.~Shapoval, {\it Pis'ma Zh.~Exp.~Teor.~Fiz.}
{\bf 64}, 570 (1996).

\bibitem{Liu1}  M.~Liu, D.~Y.~Xing, and Z.~D.~Wang, Phys. Rev. B {\bf 55},
3181 (1997).

\bibitem{Pash}  E. A.~Pashitskii and V. I.~Pentegov,
{\it Zh.~Exp.~Teor.~Fiz.} {\bf 111}, 298 (1997). [JETP {\bf 84}, 164 (1997)].

\bibitem{Beal1}  M.~T.~Beal-Monod, Phys. Rev. B {\bf 58}, 8830 (1998);
Physica C {\bf 298}, 59 (1998).

\bibitem{Sch}  I.~Sch\"{u}rrer, E.~Schachinger, and J.P.~Carbotte, Physica C
{\bf 303}, 287 (1998).

\bibitem{Modr}  R.~Modre, I.~Sch\"{u}rrer, and E.~Schachinger, Phys. Rev. B
{\bf 57}, 5496 (1998).

\bibitem{YuraLT22}  Yu. A. Nefyodov, A. A. Golubov, M. R. Trunin, and
M.~T.~Beal-Monod, to be published in Physica B.

\bibitem{Bon1}  D.~A.~Bonn, P.~Dosanjh, R.~Liang, and W.~N.~Hardy, Phys.
Rev. Lett. {\bf 68}, 2390 (1992).

\bibitem{Bon2}  K.~Zhang, D.~A.~Bonn, R.~Liang, D.~J.~Baar, and W.~N.~Hardy,
Appl. Phys. Lett. {\bf 62}, 3019 (1993).

\bibitem{Bon4}  D.~A.~Bonn, R.~Liang, T.~M.~Riseman, D.~J.~Baar,
D.~C.~Morgan, K.~Zhang, P.~Dosanjh, T.~L.~Duty, A.~MacFarlane, G.~D.~Morris,
J.~H.~Brewer, W.~N.~Hardy, C.~Kallin, and A.~J.~Berlinsky, Phys. Rev. B
{\bf 47}, 11314 (1993).

\bibitem{Kit}  H.~Kitano, T.~Shibauchi, K.~Uchinokura, A.~Maeda, H.~Asaoka,
and H.~Takei, Phys. Rev. B {\bf 51}, 1401 (1995).

\bibitem{Tru100}  M.~R.~Trunin, J.~Superconductivity {\bf 11}, 381 (1998).

\bibitem{Bul}  S.~M.~Quinlan, D.~J.~Scalapino, and N.~Bulut, Phys. Rev. B
{\bf 49}, 1470 (1994).

\bibitem{El1}  G.~M.~Eliashberg, {\it Zh.~Exp.~Teor.~Fiz.} {\bf 38}, 966
(1960) [JETP {\bf 11}, 696 (1960)]; {\it Pis'ma Zh.~Exp.~Teor.~Fiz.}
{\bf 48}, 275 (1988).

\bibitem{Pick1}  W. E. Pickett, J.~Superconductivity {\bf 4}, 397 (1991).

\bibitem{Gin}  V.~L.~Ginzburg and E.~G.~Maksimov, {\it Sverkhprovodimost':
Fiz.,~Khim.,~Tekh.} {\bf 5}, 1543 (1992).

\bibitem{Zhu2}  A.~A.~Golubov, M.~R.~Trunin, A.~A.~Zhukov, O.~V.~Dolgov, and
S.~V.~Shulga, J. Phys. I France {\bf 6}, 2275 (1996).

\bibitem{Bil}  A.~Bille and K.~Scharnberg, J. Phys. Chem. Solids {\bf 59},
2110 (1998).

\bibitem{Abra1}  C. M.~Varma, P. B.~Littlewood, S.~Schmitt-Rink,
E.~Abrahams, and A.~E.~Ruskenshtein, Phys. Rev. Lett. {\bf 63}, 1996 (1989).

\bibitem{Abra2}  E.~Abrahams, J. Phys. I France {\bf 6}, 2191 (1996).

\bibitem{PW1}  P.~W.~Anderson, {\it Physica C} {\bf 185-189}, 11 (1991).

\bibitem{PW2}  P.~W.~Anderson, {\it Theory of Superconductivity in the
High-$T_c$ Cuprates} (Princeton: Princeton University Press, 1997).

\bibitem{Plee}  P.~A.~Lee, Phys. Rev. Lett. {\bf 71}, 1887 (1993).

\bibitem{Tru3}  M.~R.~Trunin, A.~A.~Zhukov, G.~E.~Tsydynzhapov,
A.~T.~Sokolov, L.~A.~Klinkova, and N.~V.~Barkovskii, {\it Pis'ma
Zh.~Exp.~Teor.~Fiz.} {\bf 64}, 783 (1996) [JETP Lett. {\bf 64}, 832 (1996)].

\bibitem{Fink1}  H.~J.~Fink, Phys. Rev. B {\bf 58}, 9415 (1998);
H.~J.~Fink, unpublished.

\bibitem{Fink2}  H.~J.~Fink and M.~R.~Trunin, to be published in Physica B.

\bibitem{Gor}  C.~S.~Gorter and H.~Casimir, {\it Phys.~Z.} {\bf 35}, 963
(1934).

\bibitem{Tru10}  M. R. Trunin, to be published in Physica B.

\bibitem{Glas}  N.~E.~Glass and W.~F.~Hall, Phys. Rev. B {\bf 44}, 4495
(1991).

\bibitem{Ols}  H.~K.~Olsson and R.~H.~Koch, Phys. Rev. Lett. {\bf 68}, 2406
(1992).

\bibitem{Gol1}  A.~A.~Golubov, M.~R.~Trunin, S.~V.~Shulga, D.~Wehler,
J.~Dreibholz, G.~M\"{u}ller, and H.~Piel, Physica C {\bf 213}, 139 (1993).

\bibitem{Horb}  M.~L.~Horbach, W.~van~Saarlos, and D.~A.~Huse, Phys. Rev.
Lett. {\bf 67}, 3464 (1991).

\bibitem{Anl1}  S.~M.~Anlage, J.~Mao, J.~C.~Booth, D.~H.~Wu, and J.~L.~Peng,
Phys. Rev. B {\bf 53}, 2792 (1996).

\bibitem{El3}  G.~M.~Eliashberg, {\it Zh.~Exp.~Teor.~Fiz.} {\bf 39}, 1437
(1960) [JETP {\bf 12}, 1000 (1961)].

\bibitem{Mak1}  A.~E.~Karakozov, E.~G.~Maksimov, and S.~A.~Mashkov,
{\it Zh.~Exp.~Teor.~Fiz.} {\bf 68}, 1937 (1975) [JETP {\bf 41}, 971 (1976)].

\bibitem{Mar}  F.~Marsiglio, Phys. Rev. B {\bf 44,} 5373 (1991).

\bibitem{Mak2}  A.~E.~Karakozov, E.~G.~Maksimov, and A.~A.~Mikhailovskii,
{\it Zh.~Exp.~Teor.~Fiz.} {\bf 102}, 132 (1992) [JETP {\bf 75} (1), 70
(1992)].

\bibitem{Dol1}  O.~V.~Dolgov, E.~G.~Maksimov, A.~E.~Karakozov, and
A.~A.~Mikhailovsky, Solid State Comm. {\bf 89}, 827 (1994).

\bibitem{El2}  G.~V.~Klimovich, A.~V.~Rylyakov, and G.~M.~Eliashberg,
{\it Pis'ma Zh.~Exp.~Teor.~Fiz.} {\bf 53}, 381 (1991) [JETP Lett. {\bf 53},
399 (1991)].

\bibitem{Kar1}  A.~A.~Mikhailovsky, S.~V.~Shulga, A.~E.~Karakozov,
O.~V.~Dolgov, and E.~G.~Maksimov, Solid State Comm. {\bf 80}, 511 (1991).

\bibitem{Coll}  R.~T.~Collins, Z.~Schlesinger, F.~Holtzberg, C.~Field,
U.~Welp, G.~W.~Crabtree, J.~Z.~Liu, and Y.~Fang, Phys. Rev. B {\bf 43}, 3701
(1991).

\bibitem{Ram}  J.~Rammer, Europhys. Lett. {\bf 5}, 77 (1991).

\bibitem{Andre}  A.~Andreone, C.~Cantoni, A.~Cassinese, A.~Di~Chiara, and
R.~Vaglio, Phys. Rev. B {\bf 56}, 7874 (1997).

\bibitem{Maks10}  E.~G.~Maksimov, D.~Yu.~Savrasov, and S.~Yu.~Savrasov,
{\it Uspekhi Fiz.~Nauk}, {\bf 167}, 354 (1997).

\bibitem{Tru1}  M.~R.~Trunin, A.~A.~Zhukov, and A.~T.~Sokolov,
{\it Zh.~Exp.~Teor.~Fiz.} {\bf 111}, 696 (1997) [JETP {\bf 84}, 383 (1997)].

\bibitem{El4}  G.~M.~Eliashberg, G.~V.~Klimovich, and A.~V.~Rylyakov, J.
Supercond. {\bf 4}, 393 (1991).

\bibitem{Hettler}  M.~H.~Hettler and P.~J.~Hirschfeld, preprint
cond-mat/9907150, unpublished.

\bibitem{Berlin}  A.~J.~Berlinsky, D.~A.~Bonn, R.~Harris, and C.~Kallin,
preprint cond-mat/9908159, unpublished.

\bibitem{DurLee}  A.~C.~Durst and P.~ A.~Lee, preprint cond-mat/9908182,
 unpublished.

\end{references}
\end{document}